\documentclass[11pt,a4paper]{article}
\usepackage[utf8]{inputenc}
\usepackage{fullpage}
\usepackage{lastpage}
\usepackage{pdfpages}
\usepackage[colorlinks=true, linkcolor=black, citecolor=blue]{hyperref}
\usepackage{url}
\usepackage{hhline}
\usepackage{amsmath}
\usepackage{mathtools}
\usepackage{amssymb}
\usepackage{graphicx}
\usepackage{tabularx}
\usepackage{ltablex}
\usepackage{hyperref}
\usepackage{soul}
\usepackage{xcolor}
\usepackage[labelfont=bf, hypcap=false]{caption}
\usepackage{multicol}
\usepackage{ragged2e}
\usepackage{subcaption}
\usepackage{breakcites} 
\usepackage{natbib}
\usepackage{listings}
\usepackage{multicol}
\usepackage{bm}
\usepackage{soul}

\newcommand{\apj}{ApJ}

\newcommand{\apjs}{ApJ}
\newcommand{\mnras}{MNRAS}
\newcommand{\aap}{A\&A}
\newcommand{\aj}{AJ}
\newcommand{\nat}{Nature}

\newcommand{\pasa}{PASA}
\newcommand{\angstrom}{\text{\normalfont\AA}}

\usepackage[official]{eurosym}

\usepackage{fancyhdr}

\begin{document}

\begin{titlepage}

\newcommand{\HRule}{\rule{\linewidth}{0.5mm}}
\center

\textsc{\Large University of Groningen \\Kapteyn Astronomical Institute}\\[0.5cm]

\begin{figure}[h!]
 \centering
 \includegraphics[width=0.2\linewidth]{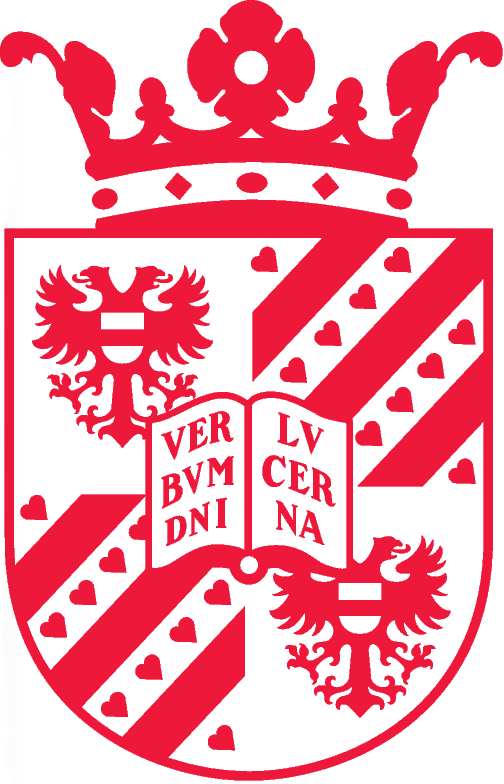}
 \label{fig:speciation}
\end{figure}

\textsc{\Large }\\[0.5cm]
\textsc{\large Master Thesis Astronomy}\\[0.5cm]

\HRule \\[0.5cm]
{ \huge \bfseries In search of the weirdest galaxies in \\the Universe}\\[0.5cm]
\HRule \\[5cm]

\begin{minipage}[t]{0.4\textwidth}
\begin{flushleft} \large
\emph{Author:}\\
Job Formsma\\
S2758717
\end{flushleft}
\end{minipage}
~
\begin{minipage}[t]{0.4\textwidth}
\begin{flushright} \large
\emph{Supervisors:}\\
Prof. dr. Reynier Peletier\\
Teymoor Saifollahi\\[0.5cm]
\end{flushright}
\end{minipage}\\[1cm]

{July 16, 2020}\\[1cm]
\clearpage
\begin{center}
  {\bf \large Abstract}
\end{center}
\justify

Weird galaxies are outliers that have either unknown or very uncommon features making them different from the normal sample. These galaxies are very interesting as they may provide new insights into current theories, or can be used to form new theories about processes in the Universe. Interesting outliers are often found by accident, but this will become increasingly more difficult with future big surveys generating an enormous amount of data. This gives the need for machine learning detection techniques to find the interesting weird objects. In this work, we inspect the galaxy spectra of the third data release of the Galaxy And Mass Assembly survey and look for the weird outlying galaxies using two different outlier detection techniques. First, we apply distance-based Unsupervised Random Forest on the galaxy spectra using the flux values as input features. Spectra with a high outlier score are inspected and divided into different categories such as blends, quasi-stellar objects, and BPT outliers. We also experiment with a reconstruction-based outlier detection method using a variational autoencoder and compare the results of the two different methods. At last, we apply dimensionality reduction techniques on the output of the methods to inspect the clustering of similar spectra. We find that both unsupervised methods extract important features from the data and can be used to find many different types of outliers.

\vfill 

\end{titlepage}

\tableofcontents

\clearpage

 
\section{Introduction}


The most interesting discoveries in Astronomy are the accidental and unintended discoveries of unknown objects never seen before. These \textit{unknown unknowns} are interesting as they can give new insights to our understanding of physical processes in the Universe, or introduce new questions about current theories in Astronomy. Finding these unknown objects is not easy as, by definition, there is no indication of what to look for. Whenever an unknown object is found, it may easily be disregarded as an error as it might look very different from the expected observations. It is up to the mind-set of the observer to either investigate the unknown finding or disregard it as an anomaly and move on with the main objective of their observations. An excellent example is the discovery of pulsars \citep{Bell_1968}, where due to good knowledge of the equipment and a persistent open mindset \citep{Norris_2017} the new type of astronomical object could be discovered. A more recent famous example is \textit{Hanny's Voorwerp} \citep{Jozsa_2009}, which was accidentally found while looking through an enormous amount of data with the citizen science project Galaxy Zoo\footnote{\url{https://www.zooniverse.org/projects/zookeeper/galaxy-zoo/}}. This resulted in an extensive search for more \textit{Voorwerpjes}, of which multiple are found nowadays giving insight into the ionization processes of Active Galactic Nuclei \citep{Keel_2012}. With the increasing size of the already big Astronomical surveys and the vast amount of data that comes from them, finding the interesting unknown unknowns by accident will become more difficult. The traditional analysis on the data surveys becomes too much work to do by hand, resulting in the need to develop new or use existing visualization techniques. 


\subsection{Machine Learning in Astronomy}
In the last decade, machine learning has been widely used in Astronomy \citep{Baron_2019}. Applications range from supervised tasks, such as classifying galaxy types, to the unsupervised tasks of clustering similar objects. \textit{Supervised learning} is used to assign a specific class for a set of data. A basic example is the classification of galaxy types based on their morphology. Information about the shape of the galaxy can be learned from pictures, where a (non-)linear relationship can be found between pixel values and galaxy types. Once the underlying relationship is found, the model can be applied to new data to predict galaxy morphology classes. Learning this relationship requires prior knowledge of the data before training, as the classes have to be known beforehand. For galaxy types, this is often done by eye by an expert in the field, or via citizen-science projects like Galaxy Zoo. Supervised machine learning can be a very strong method to automate the classification of new data as shown in \citet{Ksoll_2018}, where three different supervised methods were successfully applied to identify the Pre-Main-Sequence stars using photometric observations. \textit{Unsupervised learning} is used to learn (complex) relationships that exist in a data set. In contrast to supervised learning, there are no pre-defined classes resulting in very different applications: clustering of similar objects, dimensionality reduction, and anomaly detection. These applications are arguably more important for scientific research than classification via supervised learning, as they can be used to extract new information from a data set. For example, in \citet{Turner_2018} they applied a simple clustering method using nearest neighbors on galaxy feature data to find galaxies in their different evolutionary paths. This robust method of early data inspection can easily be scaled for large data sets. This is an excellent application of an unsupervised method, as they extracted information from a data set without the need for prior knowledge or labeled data.

\clearpage


\subsection{Outlier Detection}
One of the most interesting applications of unsupervised learning is novelty detection, or \textit{outlier detection}. \textit{Outliers} are objects that have either weird or uncommon features, making them very different from the normal sample. Many types of different algorithms can be applied to big data sets, which learn the underlying rules of common features and find objects that fall outside of these rules. Most of the outliers will consist of noisy or erroneous data, as these are simply not comparable to the real data. Outlier detection is interesting for Astronomy as it can be used to find uncommon objects, or common events that only happen on small timescales. Targeted observations of these events can be very difficult, as one does not know where to look at which time. Fortunately, the Universe is very big and with the use of big surveys, such as the Sloan Digital Sky Survey (SDSS, \citealp{SDSS_2017}) or Galaxy And Mass Assembly (GAMA, \citealp{Baldry_2017}), these rare events can still be observed by accident. \\

Next to the uncommon events and short events, a third interesting outlier category is the 'unknown unknowns'. This category consists of objects that have never been observed before and when found not, or only partly, understood. Finding these objects is exciting for Astronomers, as new information can always confirm or shake up some theories about processes and events in the Universe. Searching for 'unknown unknowns' in a data set is not a trivial task as it is a very advanced application of unsupervised learning with no indication of what to look for. Also, different outlier detection techniques may identify different objects as outliers as the methods inherently differ from each other. While each algorithm finds its own interesting outliers, there is no single method to identify all the interesting objects at once. A review of novelty detection methods is presented in \citet{Pimentel_2014}, where five different types of outlier detection techniques are described and investigated. In this work, we will apply two of these methods, distance based and reconstruction based novelty detection, on galaxy spectra to find outlying objects. \\

\textit{Distance based} outlier detection methods use an algorithm to determine distances between objects via their features or values and assign a score for their dissimilarities. An outlier score can be determined for every object using these distances, representing its difference with all other objects. The objects with the highest overall outlier scores can be considered the outliers of the data set. In \citet{Baron_2016}, they introduced a general distance based outlier detection algorithm and applied it on galaxy spectra. This resulted in the findings of many interesting outliers and their algorithm is used in our research.\\

\textit{Reconstruction based} algorithms learn the underlying rules in the data and reconstruct the data using these rules. The difference between the original input and the reconstructed data is used to compute a reconstruction error. As the model is not trained to reconstruct the unknown or uncommon features in outliers, the reconstruction error can be used as an outlier score. The application of reconstruction based methods was tested in \citet{Ichinohe_2019} and \citet{Portillo_2020}, where variational autoencoders are applied on spectra and used for outlier detection. \\

Both outlier detection methods result in an outlier score for each object, often normalized between 0 and 1. It is up to the user to determine a cut-off value, for which all objects above that value are considered to be outliers.

\clearpage


\subsection{Project Outline}
In this project, we will use two outlier detection methods to find weird interesting objects in a big astronomical survey. We try to find weird objects by looking at the spectra in the third data release of the Galaxy and Mass Assembly (GAMA) survey \citep{Baldry_2017}. Spectra contain a lot of information and physical parameters of the observed objects and are therefore very nice to use for outlier research. A description of the GAMA data and its content can be found in Section \ref{sec:data}. First, we inspect the data for possible instrumental errors, highly uncertain redshift calculations, and inaccurate flux calibrations. We apply the distance-based outlier detection algorithm developed in \citet{Baron_2016} on our GAMA data. In their work, they applied an algorithm based on Unsupervised Random Forest (URF, \citealp{Shi_2005}) on SDSS galaxy spectra and could find many outlying spectra, such as spectra with extreme line emission ratios or galaxies with supernovae. Many of their findings were not earlier reported in the literature. A comparison between this method and a few other conventional outlier detection algorithms is provided in \citet{Reis_2019}, which states that overall the URF performs the best. A thorough explanation and our implementation of the URF can be found in Section \ref{sec:urf}. In this section, we also test the algorithm using known outliers and self-generated weird spectra. The results of the algorithm are inspected and the weirdest spectra are investigated in Section \ref{sec:outlier}.\\

Next to the distance based URF method, we experiment with a reconstruction based outlier detection technique utilizing a neural network. We apply an Information Maximizing Variational Autoencoder \citep{Zhao_2017} on the galaxy spectra, which is a modified version of a variational autoencoder \citep{Kingma_2013}. This method was demonstrated on SDSS data in \citet{Portillo_2020} and can map spectra to a (very) low dimension and fully reconstruct different types of galaxies. We use the reconstruction error and low dimensional representation to determine outliers and compare the results of both methods in Section \ref{sec:outlier}. \\

Additional to the search for outliers, we also inspect the high dimensional output of the outlier detection methods using t-SNE \citep{Maaten_2008}, a visualization technique used to reduce high-dimensional data to a two or three-dimensional map. This was also done in \citet{Reis_2018a}, where they applied t-SNE on the output of the URF algorithm. They showed that the high dimensional output contains a lot of complex information of the data. With the map, the structure of the data can be projected and used to group and find similar galaxies. We show the application of t-SNE and the maps in Section \ref{sec:tsne}. \\

The main objective of this research is to find the outliers and weirdest objects in the GAMA survey, but we also want to promote the use of outlier detection methods on spectroscopic data. Spectra are very interesting to look at as they trace many physical properties of the object. In the near future, new surveys like WEAVE \citep{Dalton_2014} and 4MOST \citep{4MOST_2019} will generate an enormous amount of spectra. Finding interesting outliers or anomalies becomes more difficult as more spectra have to be inspected. To convince people for citizen science projects might be difficult, as spectra do not look as interesting as the sky pictures used in the citizen science projects like GalaxyZoo. To cope with the amount of data we need good machine learning techniques to automatically find the interesting outliers in future surveys. We use and compare two outlier detection algorithms and explore their usability for the search of outliers in the Universe. 

\clearpage
\section{Data}\label{sec:data}
We use data from the third data release (DR3) of the Galaxy and Mass Assembly (GAMA) survey \citep{Baldry_2017}. The GAMA survey is a spectroscopic and multiwavelength photometric survey, with the main objective to study cosmic structures on a scale of 1 kpc to 1 Mpc \citep{Driver_2009}. This includes the study of galaxy clusters, groups, and mergers to test the cold dark matter paradigm of structure formation. The third data release consists of over 150000 spectra of objects with a reliable heliocentric redshift \citep{Baldry_2014} and other additional photometric parameters \citep{Hill_2011}. The objects are located in three equatorial and two southern observational regions covering a total area of 286 deg$^2$. Primary observations were done using the AAOmega spectrograph at the Anglo-Australian Telescope \citep{Smith_2004}. \\

In total, the GAMA survey consists of more spectra than the observed spectra of the AAOmega spectrograph. Before the survey started an input catalog was assembled of the objects in the observation regions to make a highly complete redshift survey. The object locations were drawn from the SDSS DR6 photometric observations and UKIDDS \citep{Lawrence_2007}, with magnitude limits of $r < 19.5$, $z < 18.2$ and $K_\text{AB} < 17.6$ \citep{Baldry_2010}. This is a very high-density sample of the objects in the observational regions and therefore a nice survey to use for outlier detection. Most objects in the observation regions overlap with earlier spectroscopic surveys, such as SDSS, the 2dF Galaxy Redshift Survey \citep{Colless_2003}, and Millennium Galaxy Catalogue \citep{Driver_2017}. As a result, if an object is observed in different surveys, the best spectrum can be chosen to get a complete data set with good spectra. In this project, we only use the spectra from the main observations with the AAOmega instrument, which has a spectral resolution of $R \approx 1300$. These spectra make up over 78\% of the total spectra and is thus a sufficient representation of the survey, while also making it easy for us to only have to work with data from a single instrument. In comparison with other big surveys, the GAMA survey contains high-resolution spectra of fainter objects at a higher median redshift. This makes it an interesting data set for this research, as we can search for outliers deeper in space.\\

The data used in this project is obtained via the GAMA DR3 website\footnote{\url{http://www.gama-survey.org/dr3/}}. We select the spectra by setting the \texttt{SURVEY\_CODE} to 5, which represents spectra from the main observations, and \texttt{IS\_SBEST} to be true, such that we only have the best spectrum for every single object. Another important parameter is the \texttt{nQ} value, which represents the normalized quality factor of the redshift value fitted to the spectra. For quality research, we only use spectra with a value of $\text{nQ} >= 3$, as advised in \citet{Hopkins_2013}. The full data query can be found in Appendix A, which gives a total of usable spectra to be just over 130000. The wavelength range of the instrument, and therefore the observed wavelength range of the objects, is between 3750 \angstrom\ and 8850 \angstrom\ over 4954 pixels. 

\clearpage
\begin{figure}[ht!]
    \centering
    \includegraphics[width=\linewidth]{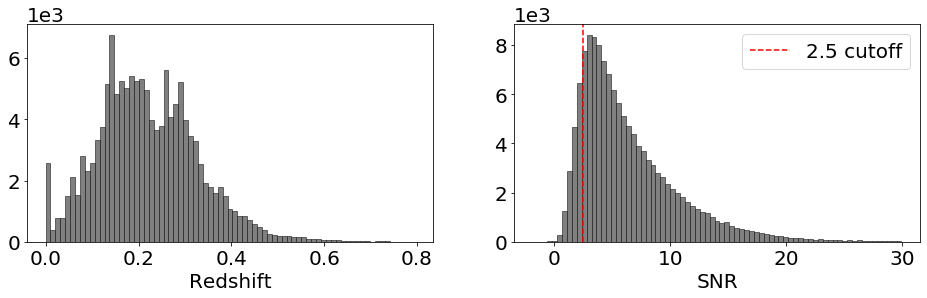}
    \caption{Overview of distributions of GAMA parameters}
    \label{fig:inspection}
\end{figure}
\subsection{Inspection}
There is a vast amount of different spectra in the GAMA survey. We quickly inspect some parameters that are already provided for each spectrum: a fitted redshift parameter and a signal to noise ratio SNR. The distributions of these parameters are shown in Figure \ref{fig:inspection}. Looking at the redshift distribution we can see that the median redshift of GAMA objects is $z=0.21$. We also see two big global peaks and a few smaller local peaks in the histogram. The two big peaks follow from clusters in the observed fields, while the smaller peaks represent clustering on smaller scales. Note that there is also a peak at a redshift of $z=0$. Objects with this assigned redshift value are: foreground stars, point-like quasar potentials which turned out to be stars, galaxy-star blends, and bad spectra with a wrong fitted redshift value. As a test, we run these objects through our reconstruction based clustering algorithm. This resulted in the classification of 1567 M-dwarfs, 144 O-stars, and 1118 stars that are either a type A, F or G-star. These spectra will be left out of the final classification of galactic outliers in Section \ref{sec:outlier}.

\subsection{Pre-processing}\label{sec:preprocessing}

\begin{figure}[ht!]
    \centering
    \includegraphics[width=\linewidth]{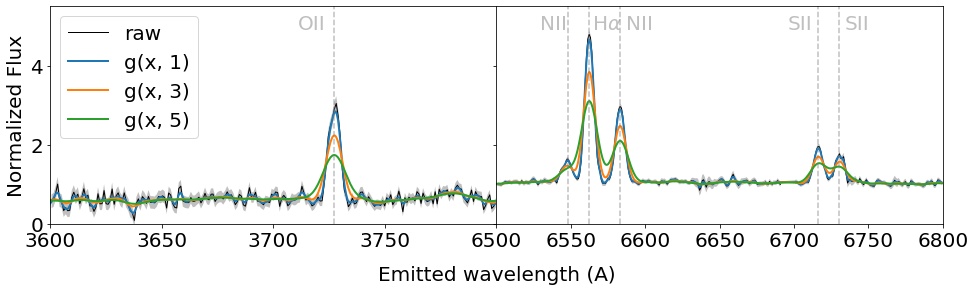}
    \caption{Different Gaussian kernel widths tested. The labels of the smoothed spectra refer to the Gaussian kernel found in \eqref{eq:kernel}.}
    \label{fig:kernels}
\end{figure}

In this project, we use the flux values of the spectra as the input of the outlier detection algorithms to compute an outlier score for each spectrum indicating its weirdness. The output of the algorithms is thus entirely based on the shape of, and features in, the spectrum. To get the best unbiased results, all the raw flux values could be used as input for the outlier detection algorithm. However, the raw data contains a lot of bad data points, or noisy continuum as shown in Figure \ref{fig:kernels}, which interfere with the results. A few bad pixels, for example due to a cosmic ray, can cause a high outlier score for an otherwise normal spectrum. We apply a few pre-processing steps on the raw data to make it suitable for our project. We try to keep the processed data as close to the raw data as possible, as this will give the best relation between an outlier score and the observed spectrum.\\

First, we mask three noisy regions in all spectra. The blue end of the spectra up to the observed wavelength of 4050 \angstrom\ is masked due to low flux values as a result of high sky absorption. We also mask the region between 5570 \angstrom\ and 5585 \angstrom\ due to high residual sky emission seen in most of the spectra. At last, the region with wavelengths larger than 8780 \angstrom\ is masked as there are noisy points at the end of most spectra. Other regions with bad pixels, which are not at the end of the spectra, are linearly interpolated, to simulate as if those regions are normal. 

\clearpage

Next to removing the bad pixels, we smooth the data by applying a low pass filter. The raw spectra show a lot of flux fluctuations in the parts where there are no emission lines, especially in the blue end of the spectra. Due to the randomness of these fluctuations, spectra can be assigned a high outlier score without the presence of real physical features. We smooth the flux values in the spectra via convolution with the Gaussian kernel using the Astropy\footnote{\url{https://docs.astropy.org/en/stable/convolution/}} package in Python. The kernel can be described as
\begin{equation}\label{eq:kernel}
    g(x, \sigma) = \frac{1}{\sqrt{2\pi\sigma^2}}e^{-\frac{x^2}{2\sigma^2}}\ ,
\end{equation}
where $\sigma$ denotes the standard deviation in terms of pixels. This kernel suppresses the low-level fluctuations, while keeping the shape of the emission lines intact. We test with different kernel sizes, to find an optimal balance between suppressing noise and keeping the data close to the original. This is seen in Figure \ref{fig:kernels}, where the kernels with different pixel widths are applied on the raw spectra and shown on top of each other. The important tasks are keeping the shape of the emission lines as authentic as possible, while reducing the fluctuations on the continuum level. We see that the Gaussian kernel $g(x, 3)$ is a good trade-off, but we will also run tests with smaller and bigger kernel widths.\\

Redshift is a non-linear variable, so we want to avoid working in the observed frame. We redshift the spectra to their rest-frame wavelengths and interpolate all spectra to a common grid. As flux values will be used as feature input for the outlier detection algorithms, it is important to have all emission lines at the same input point. We interpolate all spectra between the rest-frame wavelengths of 3500\angstrom\ and 7500\angstrom to 8000 data points, giving a slightly over sampled resolution of 0.5\angstrom\ per data point. This is to ensure that there are enough data points to probe the width and separation of emission lines for all redshifted spectra. Missing flux values at the ends of the spectra are extrapolated as a flat line with the average values of nearby points. At last, the full spectrum is normalized by dividing by the median of the flux values. We use three different subsets in this project based on the \textit{signal to noise ratio} (SNR) of the spectra. An overview of the subsets is shown in Table \ref{tab:subsets}. 

\vspace{1cm}

\clearpage

\subsection{Bad Spectra}\label{sec:databad}
\begin{figure}[ht!]
    \centering
    \includegraphics[width=\linewidth]{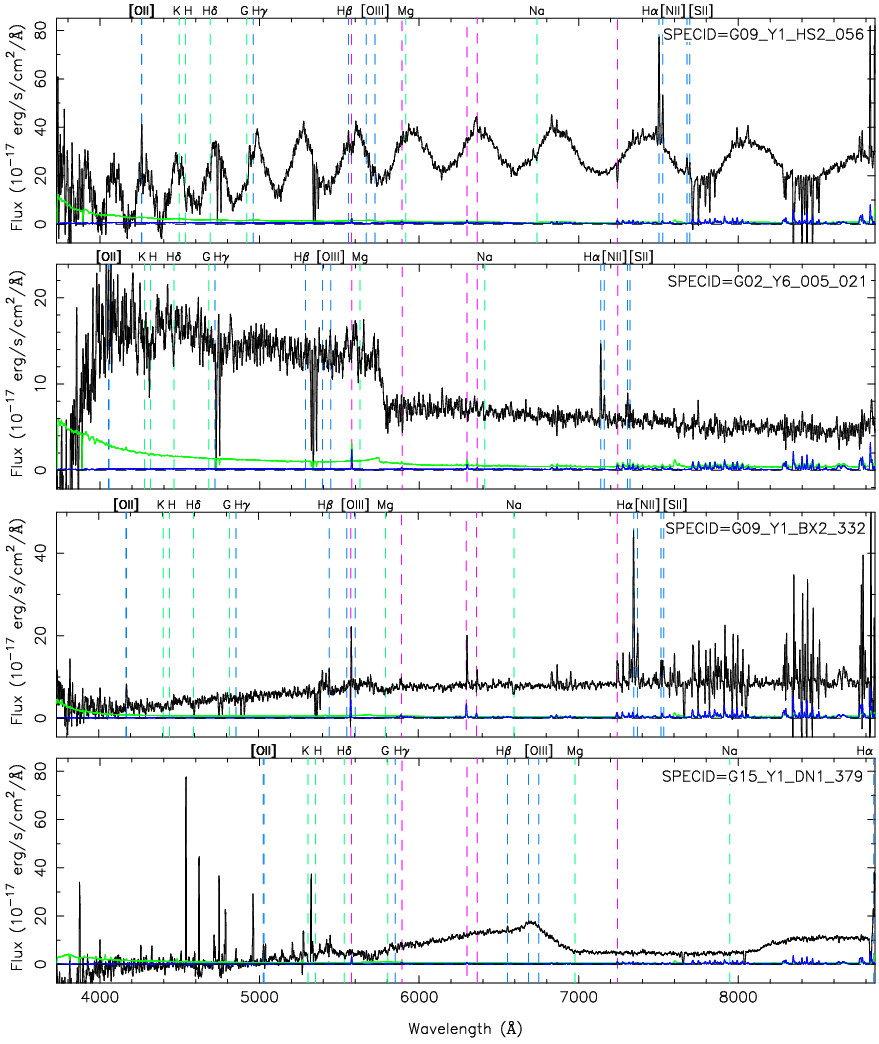} 
    \caption[ding]{Groups of bad spectra in GAMA survey. From top to bottom: fringed 
    spectra, bad splice between instrument arms, sky emission,
    DN1 field error. Pictures taken from the GAMA DR3 Single Object Viewer\footnotemark}
    \label{fig:badspectra}
\end{figure}

During early inspection of the data, a few weird spectra were commonly found. These spectra are shown in Figure \ref{fig:badspectra} and trace four of the most recurring types of bad spectra. At first, we have many fringed spectra shown by the top spectrum of Figure \ref{fig:badspectra}. These sine-like oscillations arise from shortcomings in the AAOmega instrument and are present in observations of a few specific fibers. The fringing arises from air gaps in the glue at the connection point of the fibers and the prism causing high-frequency oscillations and poor removal of sky features \citep{Hopkins_2013}. For most fringed spectra, the redshift could still be determined as emission lines are still visible. However, as the continuum clearly does not trace the physical properties of a galaxy, we exclude these fringed galaxies from our outlier detection. As the fringed galaxies have different shapes of oscillations, there is also no robust way to "defringe" all galaxies in a fully automated manner. \\

The next three classes of bad spectra are solely due to reduction errors. The second spectrum of Figure \ref{fig:badspectra} shows a very big change in continuum at 5700 \angstrom. The left and right parts of the spectra are observed by different arms of the instrument and have to be normalized to match each other. Several spectra have this so-called \textit{bad splice}, which is due to poor flat fielding in one of the arms\cite{Hopkins_2013}. Just as with the fringed spectra, a reliable redshift can be determined as emission and absorption features are visible. 
The third spectrum of Figure \ref{fig:badspectra} shows obvious sky emission lines have not been subtracted correctly. \\

The last group of bad spectra is an unexplained amount of spurious lines in a small observing region in the sky. Also, a big continuum error between 6000 \angstrom\ and 7000 \angstrom\ can be seen. This effect is not earlier reported and there seems to be no information about the source of the weird features. This type of bad spectra is found in the specific observed region \texttt{G15\_Y1\_DN1\_nnn}, with \texttt{G15\_Y1} the observed field, \texttt{DN1} the field identifier and \texttt{nnn} the fiber identifier. As each fiber in this observed field shows the same errors we call this group the \textit{DN1 field error}.\\

Initial random inspection only gave us a few spectra of each group. However, as these spectra are very different than normal spectra, the outlier detection algorithms naturally picked up these spectra as outliers. By iterative application of the algorithm and removal of the bad spectra, we found all spectra belonging to the above groups. In total, we found that 4.5\% of the GAMA spectra are either affected by fringing, bad splicing, or other reduction errors. In the GAMA spectroscopic analysis paper \citet{Hopkins_2013} the amount of bad spectra was reported to be around 3\%, which shows that we have found a few more than earlier reported. Furthermore, we find a total of 150 bad \texttt{DN1} field errors. \\

\begin{table}[ht!]
    \centering
    \begin{tabular}{l|l|r|r|r|r}
         \textbf{Name} & \textbf{Subset} & \textbf{Objects} & \textbf{GG Spectra} & med\{\textbf{SNR}\} & med\{\textbf{z}\}\\
         \hline
         SNR2.5          & SNR $\geq 2.5$   & 115170  & 106093 &  5.88 & 0.204\\
         SNR5            & SNR $\geq 5$     & 69560   & 63893  &  8.20 & 0.183\\
         SNR10           & SNR $\geq 10$    & 23120   & 21226  & 12.98 & 0.158
    \end{tabular}
    \caption{Subsets of GAMA AAOmega spectra. The \textit{GG spectra} indicate the size of the subsets with only good galactic spectra, excluding stars and bad spectra.}
    \label{tab:subsets}
\end{table}

\clearpage

\footnotetext{\url{http://www.gama-survey.org/dr3/tools/sov.php}}

\clearpage
\section{Outlier Detection Methods}\label{sec:urf}
We apply two different algorithms on the GAMA spectra and try to find outliers. The first algorithm we use is a distance-based algorithm using Unsupervised Random Forest (URF, \citealp{Shi_2005}), which was developed in \citet{Baron_2016} and used to detect outliers in SDSS spectra. The behavior of the algorithm is inspected via different tests to determine its ability to detect outliers and its sensitivity to noisy data. We will show the used data sets and explain the application of the algorithm on the data. We also use a reconstruction based algorithm using an Information Maximizing Variational Autoencoder (InfoVAE, \citealp{Zhao_2017}), which was used as an experiment in \citet{Portillo_2020} to synthesize SDSS spectra. We apply this algorithm as an experiment on GAMA spectra to find outliers, spectra which can not correctly be reconstructed, and to learn about the application of neural networks on astronomical data. In this section, we only look at the application process and outputs of the algorithms. The outliers of the GAMA survey are discussed in the next section.

\subsection{Unsupervised Random Forest}
\begin{figure}[ht!]
    \centering
    \includegraphics[width=\linewidth]{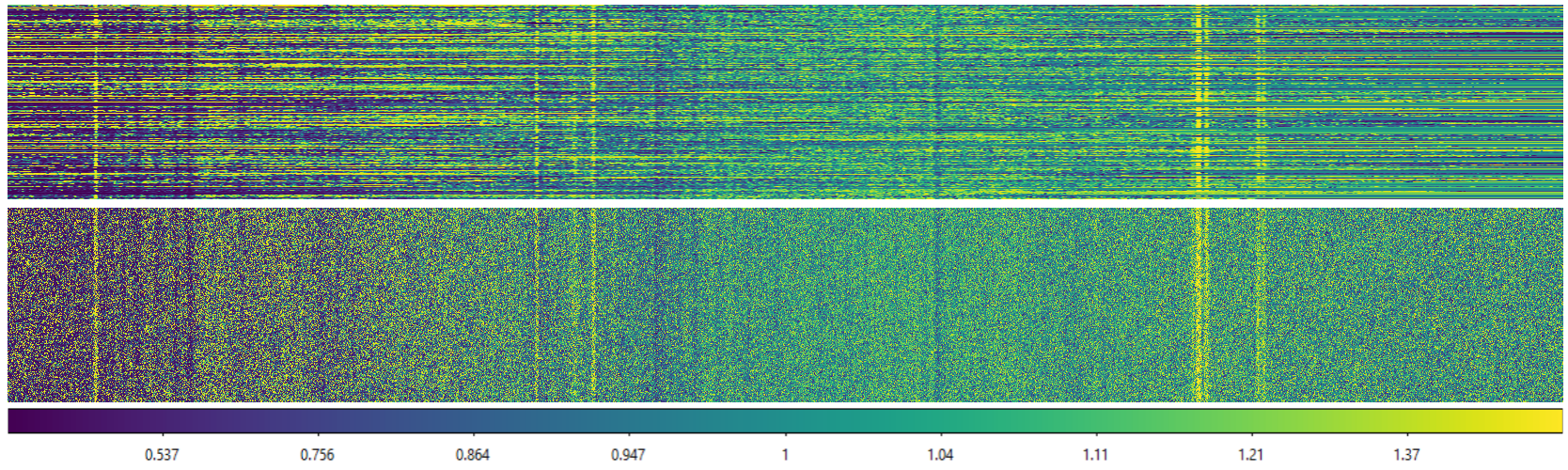}
    \caption{Comparison between real data and synthetic data. 1000 spectra
    are stacked from top to bottom with the input data points from 
    left to right. The bottom picture is synthetic data generated from the
    top picture. Normalized flux values are displayed by the color map.}
    \label{fig:realfake}
\end{figure}

The URF is used to generate the pair-wise distances between all objects by looking at the features in the data, constructing a distance matrix. As shown in \citet{Baron_2016}, this distance matrix traces a lot of information of the objects and can be used to determine an overall distance (or \textit{weirdness}) score for each object. The fundamental part of the algorithm is the Random Forest (RF, \citealp{Breiman_2001}). An RF is normally used as a supervised application, in which an ensemble of decisions trees learn the rules in the data and determine the class labels via a majority vote. Our data is not labeled, so we use the RF in an unsupervised way as described in \citet{Shi_2005}. Instead of assigning different labels to the input data, we generate synthetic spectra based on our real data and let the RF learn the difference between the real spectra and the synthesized spectra. The RF is trained with a combination of the real and synthetic data to learn the covariance in the real data and the importance of different features. The synthetic data is generated by random sampling of each feature over all spectra. This gives spectra-like shapes representing the important features in the data, while also including minor details. A comparison between the real spectra and the generated synthetic spectra is shown in Figure \ref{fig:realfake}. Some aligned features as Hydrogen and Oxygen emission lines can be traced in both the real and synthetic data.  

\clearpage
With the real and synthetic spectra, we prepare multiple chunks of 10000 objects via random sampling without replacement to train the Random Forest. In each chunk, the objects are unique, while they can still be included in multiple chunks. We train 200 decision trees on each chunk and combine these to make up the full Random Forest. This approach improves the quality and the computational time of the algorithm, as we train multiple times on small amounts of data. The number of features used in the trees is the square root of the size of the input data, which was also used in \citet{Baron_2016}.\\

\begin{figure}[ht!]
    \centering
    \includegraphics[width=0.95\linewidth]{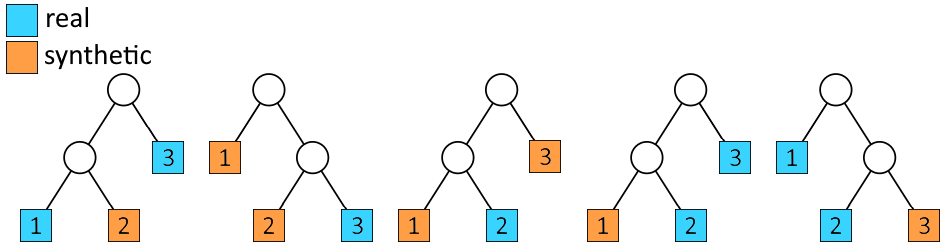}
    \caption{Very basic Random Forest with 5 decision trees. The applied input propagates through the decision nodes and ends up in a terminal leaf indicating whether it is either real or fake. An identifying number is assigned to each terminal leaf in each decision tree.}
    \label{fig:rf}
\end{figure}

After training, only the real spectra are applied to the RF to trace their similarities. The spectra propagate through all the decision trees and eventually end up in a terminal leaf, normally indicating an assigned label in supervised applications. In our application of RF, the terminal nodes indicate if the input was either real or synthetic. This is visualized in Figure \ref{fig:rf}, where a small Random Forest is shown composed of only five decision trees. The terminal leaves have an identifying number in each decision tree, and a pair-wise similarity between two spectra can be determined by looking at how often they both end up in the same terminal leaf. This value ranges from zero, indicating no similarities, to a maximum value of the number of decision trees, indicating total similarity. This value is normalized by the number of trees, and the distance score between the two objects is defined by subtracting 1 with the similarity score. This yields a distance score between 0 and 1 for every pair of objects, giving a distance matrix. The distance matrix contains information about the clustering of the spectra and can be used to trace a single \textit{outlier score} for each spectrum by averaging over the dissimilarities.\\

A small improvement of the URF proposed and used in \cite{Baron_2016} is to only count the terminal leaves of trees that label the spectra correctly as \textit{real}. The decision trees that label real spectra as synthetic for either of the two spectra in the pair-wise comparison should be excluded from the similarity comparison, as they apparently can not be trusted to identify the real features of the spectra. A very detailed description of the URF algorithm and a nice example in 2D space is shown in \citet{Baron_2016}, giving an excellent visual insight of the application of the algorithm. We implement our own code for the URF, based on the example code of the 2D example. It is fully written in Python using the scikit-learn\footnote{\url{https://scikit-learn.org/stable/modules/generated/sklearn.ensemble.RandomForestClassifier}} Random Forest implementation and is run on a 64 core computer. The full algorithm can be found on our Github page\footnote{\url{https://github.com/Formsma/GAMA-outliers}}.

\vfill
\clearpage

Primarily, we want to find outliers in the GAMA data, but we also want to investigate the results of the URF algorithm to try to understand how the distance scores are determined. For the algorithm tests, we use the high quality \texttt{SNR10} subset with 23120 objects. After the synthetic data generation the amount of objects is doubled, giving a data input of (46240 x 8000) pixels. As stated in \citet{Baron_2016}, dimensionality reduction of this input does not increase the quality of the outlier results. The first runs of the URF algorithm resulted in a lot of bad spectra and not very obvious outliers, so we want to explore how the URF works. Unsupervised machine learning can often end up in stirring the algorithm parameters until the result is satisfactory, but we prefer to explore how the algorithm works and reacts to different kinds of spectra. \\

To test the response of the algorithm, and therefore the outlier scores of the different spectra, we apply a few different versions of our data set. We first try different levels of convolution and see if the quality of the results are correlated with the amount of smoothing. Afterwards, we test different input sizes for the features of the spectra by changing the amount of interpolation. These tests only show how the URF responds to the data structure, but we also want to test the ability to find outliers. We apply the GAMA data and look in the results for the already known weird spectra reported in the GAMA database. Also, we generate \textit{fake} spectra, which are inserted into our data set acting as outliers. Note that these fake spectra are different from the synthetic spectra shown in Figure \ref{fig:realfake}, as the fake spectra are composed of our own invented shapes and the synthetic spectra are derived from random sampling of our data set.

\subsubsection{Format Tests}
\begin{figure}[ht!]
    \centering
    \includegraphics[width=0.95\linewidth]{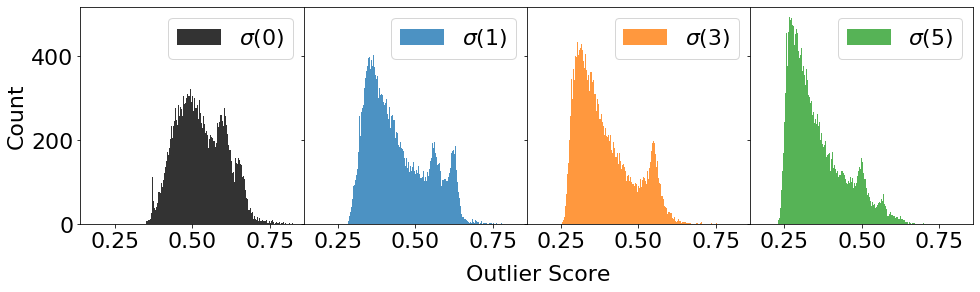}
    \caption{Distributions of different kernels used for smoothing.}
    \label{fig:kernelcompare}
\end{figure}

As discussed in Section \ref{sec:data}, the smoothing of the data is important to suppress noise on the continuum level, while keeping the emission line shapes intact. We smooth the data via convolution using a Gaussian kernel and apply the different kernel sizes shown in Figure \ref{fig:kernels} using Equation \eqref{eq:kernel}. We apply the kernels on the \texttt{SNR10} subset and use the data set on the URF algorithm. The distributions of the outlier scores are shown in Figure \ref{fig:kernelcompare}. In terms of differences, we can see that the higher the kernel, the smoother the score distribution and the lower the average score. This is very logical as smoothing removes noisy fluctuations in the data giving more similar continuum shapes. In every histogram, the main peak with the highest number of scores traces the most similar galaxies. However, we also observe one or two smaller peaks inside the distribution. After inspection, we can conclude that these smaller peaks trace the clustering of very blue and red galaxies, clustered mostly based on their continuum shape. For every distribution, except the kernel $g(x, 3)$, the individual peaks trace either the very blue or very red galaxies. For the $g(x, 3)$ kernel both the blue and red galaxies are found in the single peak. This will be more useful for our search for outliers, as these spectra are mixed and ordered on other features instead of only the continuum. With these results and the knowledge that the $g(x, 3)$ kernel preserves the emission and absorption line structure as discussed in \ref{sec:preprocessing}, we decide to use this kernel in all future runs of the algorithm.

\clearpage

The second data structure test is the variation in the input size of data into the RF. This is altered via the interpolation done on the original spectra. The more data points there are, the better the width of features can be determined. The URF does not know that emission lines can be fitted with a Gaussian, so having many data points per emission line lets the URF learn about its width. During the project various amounts of data points are used, where we eventually settled on a total of 8000. This gives around 20 data points for a strong H$\alpha$ line, while more subtle lines have at least 10 data points. By lowering the amount of data points we can not trace the width of emission and absorption features accurately. Increasing the number of data points results in increased storage space and computational time, while not increasing the quality of the results as most points are interpolated from the observations. \\

With the set number of input data points, we also comment on the number of data points used in the decision trees. Not all 8000 data points are used during learning, but a selection is made of data points in each decision tree. Only data points with a high \textit{gain} are used, which describes the amount of information that can be learned from each data point. The selection is done by taking the square root of the input as the number of decision points. This is the most common method when using Random Forests, and was also used in \citet{Baron_2016}. In our tests with varying the number of features used for decisions, we observe that for a higher than the default number of data points results in clustering based on the continuum. Keeping the number of features to be the square root of the input size ensures that the algorithm learns the best covariance of the data while we use a minimum amount of points and avoid overfitting.\\

\begin{figure}[ht!]
    \centering
    \includegraphics[width=\linewidth]{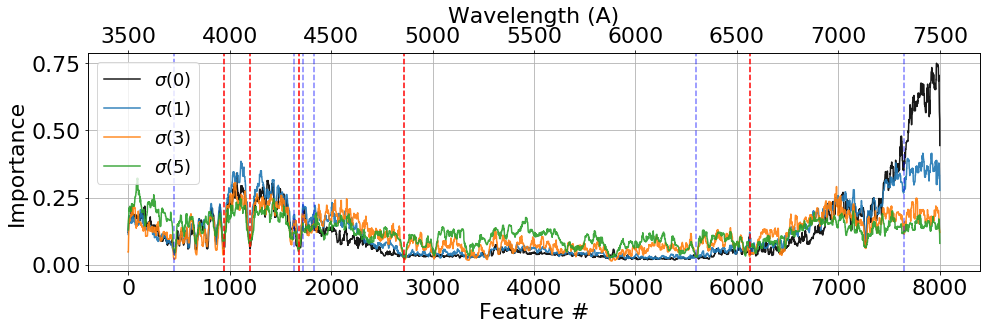}
    \caption{Relative feature importance of input features indicating which data points are used first. Hydrogen emission lines are indicated by red lines, Oxygen emission lines are indicated by blue lines.}
    \label{fig:importance}
\end{figure}

After training the URF, an interesting question is: which of the data points are used as features in the random forest and how important are the flux values at specific wavelengths? As the URF consists of many decision trees, we can not simply display all the splits in the trees in a figure. However, we can display the \textit{feature importance} of each data point and get a simple view of the usage of the features. In Figure \ref{fig:importance} the normalized feature importance is shown for the input data points. High values indicate a high entropy split and are used as first splits in the trees, while low entropy splits are used as final deciders of the class. We can see that on a continuum level the entropy is highest and for the well-known emission and absorption lines the importance is lower. This overview shows that the initial decision is made on the continuum of the spectra and the spectral lines decide the final classification. For the lower kernels, the importance is higher at the red end of the spectra. This is probably due to the differences in the redshift of the spectra where the ends of the spectra can have sharp cutoffs. This suggests a correlation between outlier scores and redshift, which could be observed very marginally.

\clearpage

The last test we perform is the training with the spectra divided over \textit{signal to noise} (SNR) bins. As most spectra have a relatively low SNR (see Fig. \ref{fig:inspection}), the URF will mostly train on those spectra. This gives that the weak emission lines present in higher quality spectra will not be used as an important feature and important information is lost. Also, a noisy continuum of a low signal to noise ratio spectrum will have a big influence on its outlier score. By dividing the spectra up into bins and train the URF on each individual bin we expect to have better results in terms of important features. This method was suggested and used in \citet{Baron_2016} to reduce the influence of low SNR spectra.\\

For the \texttt{SNR10} subset, no difference was found in dividing into bins as all the spectra were already of good quality. For the \texttt{SNR5} and \texttt{SNR2.5} subset, the training was done including bins as a correlation between the SNR and weirdness score could be observed. However, even with the binning of the spectra on their SNR, the URF algorithm is still biased towards the low SNR spectra. This was also found in \citet{Reis_2019}, where different algorithms are compared. Nonetheless, even with a bias towards low SNR spectra, the algorithm can still find weird spectra among the high SNR spectra as can be seen in Figure \ref{fig:rankedsnr}. The ranked distribution shows that weird galaxies are also found among the high SNR spectra.

\subsubsection{Searching for the Known}
The last tests we perform is looking at already known and some self-made outliers. As outlier detection is searching for the unknown unknowns, we can not directly see if the algorithm is producing good results using the GAMA data. By inspecting the already known outliers and inserting spectra (or spectra-like) shapes into the data set we can trace the outcome of different features in the data.\\

In the GAMA database, there is already a \texttt{COMMENT} flag added to some spectra with a note or remark. These comments were added during the redshift fitting and random inspection of the spectra by the GAMA team. They trace some bad spectra, but also mention some weird spectra shapes and physical events as AGNs. We apply the URF on the \texttt{SNR5} subset and overlay the known fringed spectra, bad splices, and other flagged spectra on top of the results. 

\clearpage

\begin{figure}[ht!]
    \centering
    \includegraphics[width=\linewidth]{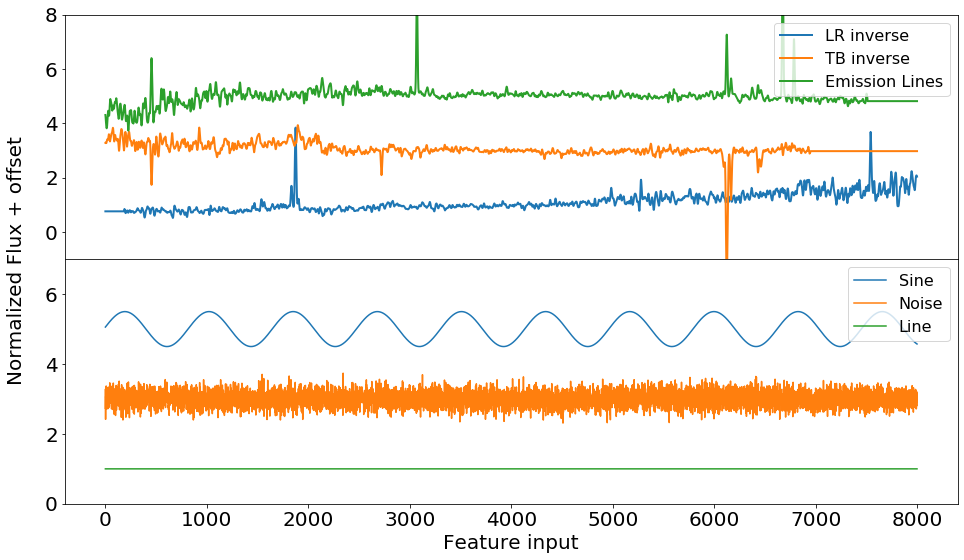}
    \caption{Types of self-made fake spectra}
    \label{fig:fakedata}
\end{figure}

We also make six types of fake spectra, divided into two classes. The first class is composed of real spectra which are transformed, as seen in the top three spectra of Figure \ref{fig:fakedata}. Spectra are flipped from left to right (LR) by rotating around the central feature, resulting in that the most right feature becomes the most left feature. Also, some spectra are flipped top to bottom (TB) rotating along the horizontal axes, which results in that emission features become absorption features. At last, we insert Gaussian features in some spectra, representing extra fake emission lines. These three types can trace the sensitivity of the URF to either the total continuum or emission lines. The second class of fake data has nothing to do with spectra, but merely are some extreme cases we put in to see how the algorithm responds. The first type is a sine wave, closely resembling the fringed spectra of the data set. The other two spectra are random noise and a flat line. We generate 50 spectra of each type in both the classes resulting in a total of 300 fake spectra.\\

\begin{figure}[ht!]
\begin{subfigure}{0.49\linewidth}
    \centering
    \includegraphics[width=\linewidth]{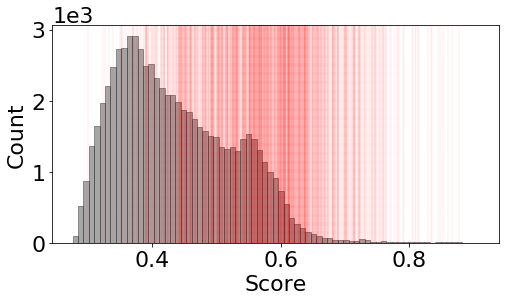}
    \caption{Fringed (from GAMA)}
\end{subfigure}
\begin{subfigure}{0.49\linewidth}
    \centering
    \includegraphics[width=\linewidth]{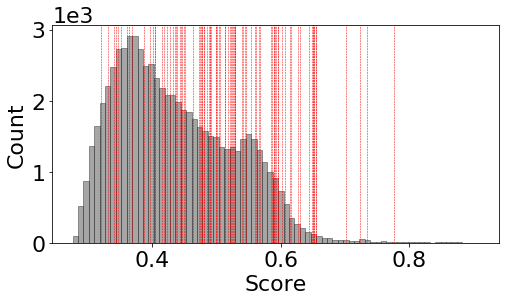}
    \caption{Reduction Error (from GAMA)}
\end{subfigure}
\begin{subfigure}{0.49\linewidth}
    \centering
    \includegraphics[width=\linewidth]{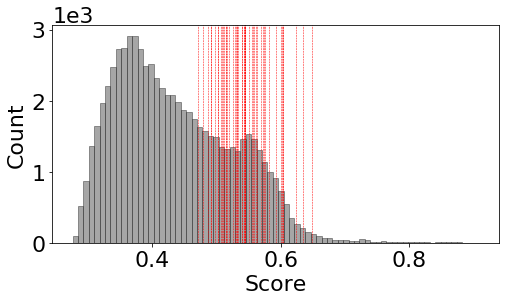}
    \caption{Left-right flip}
\end{subfigure}
\begin{subfigure}{0.49\linewidth}
    \centering
    \includegraphics[width=\linewidth]{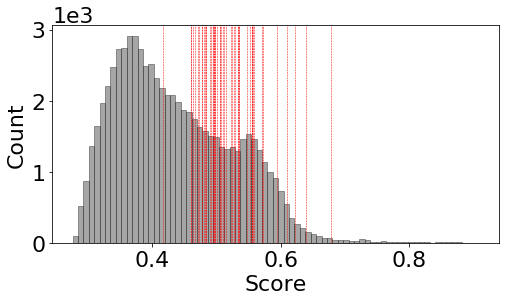}
    \caption{Top-bottom flip}
\end{subfigure}
\begin{subfigure}{0.49\linewidth}
    \centering
    \includegraphics[width=\linewidth]{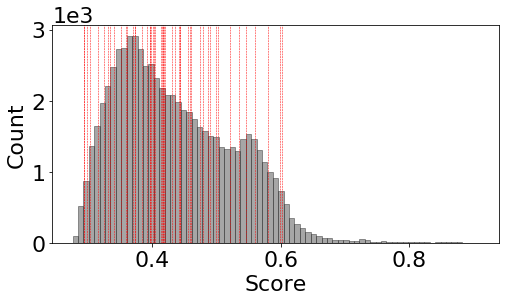}
    \caption{Extra emission lines}
\end{subfigure}
\begin{subfigure}{0.49\linewidth}
    \centering
    \includegraphics[width=\linewidth]{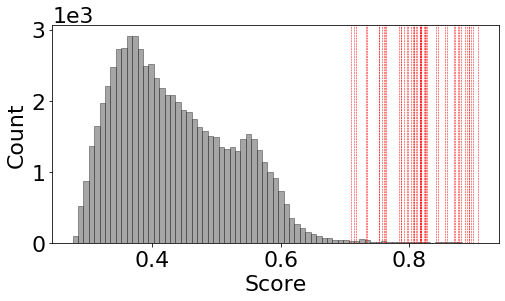}
    \caption{Sine}
\end{subfigure}
\begin{subfigure}{0.49\linewidth}
    \centering
    \includegraphics[width=\linewidth]{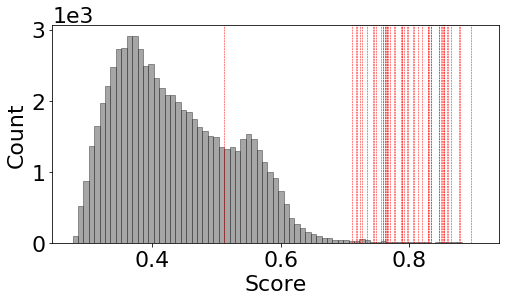}
    \caption{Random noise}
\end{subfigure}
\begin{subfigure}{0.49\linewidth}
    \centering
    \includegraphics[width=\linewidth]{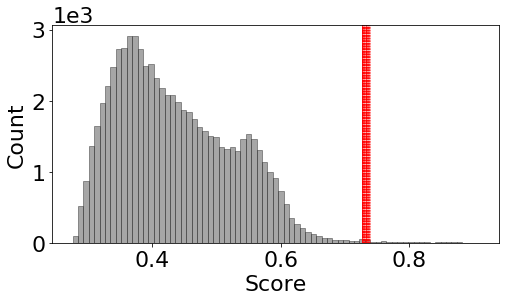}
    \caption{Single line}
\end{subfigure}
    \caption{Score distribution of the objects in the \texttt{SNR5} subset. The location of the known outliers and fake data are indicated by red lines. Known outliers in (a) and (b) are taken from the GAMA comments. For the rest of the plots, the types of fake data are shown in Figure \ref{fig:fakedata}.}
    \label{fig:known}
\end{figure}

The score of all flagged spectra, from the GAMA team and our self-made ones, are shown on the score distributions of the \texttt{SNR5} subset in Figure \ref{fig:known}. Overall we see that the outliers are all assigned a high score, except for the added emission lines. This difference is probably due to the randomness of the location of these extra emission lines and the fact that the other outliers are very weird based on their continuum. As the URF was trained on all the extreme outliers in the data set, the trained network has not learned all the real important features of the data. A run without these fake spectra would give better results, but this test already showed that the weird galaxies can be found. 

\vfill

\clearpage

\clearpage

\subsubsection{Searching for the Unknown}
To ensure real physical outlier results, the important step is to remove the bad spectra from the data set, as these are natural outliers and clutter the real interesting results. This also gives that the algorithm does not have to learn from those outliers, giving that the Random Forest is only trained on real features. During the project we flagged the spectra that belong to the groups of Section \ref{sec:databad} and also remove spectra with a redshift $z < 0.002$, representing the star-like objects. In total, we remove 8045 spectra from our subsets before applying the algorithm. We apply the algorithm on each subset of Table \ref{tab:subsets} and show the output in the following figures. We expect to get the best results with the high SNR subsets, as the quality of spectra is good. But, to ensure we find the weirdest objects of the whole data set we also run on the lower signal to noise ratio spectra.\\

\begin{figure}[ht!]
    \centering
    \includegraphics[width=\linewidth]{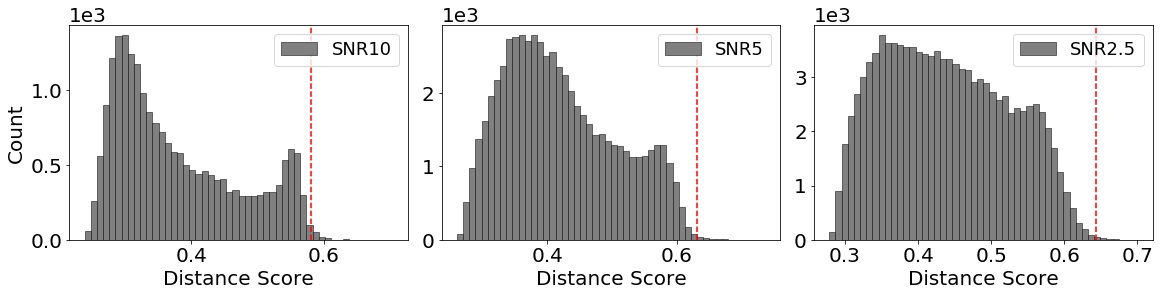}
    \caption{Score distributions of objects in the different subgroups. The 100 weirdest objects are found on the right size of the red line.}
    \label{fig:histscores}
\end{figure}

\begin{figure}[ht]
    \centering
    \includegraphics[width=\linewidth]{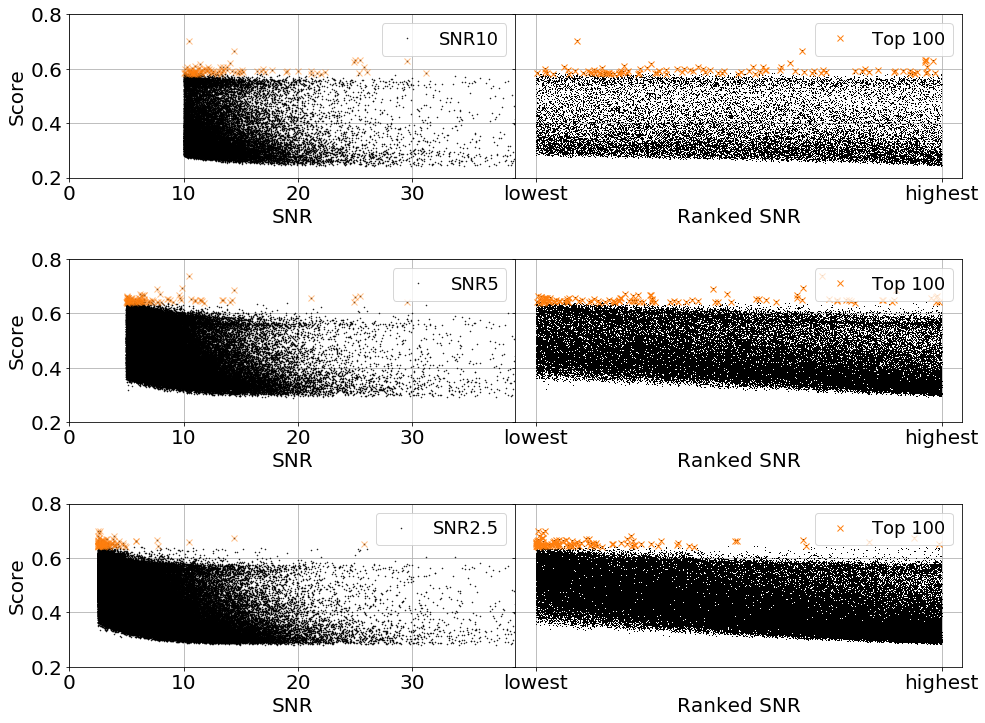}
    \caption{Relation between the distance score from the URF and the signal to noise ratio of the objects for all subsets. From top to bottom: SNR10, SNR5 and SNR2.5. Note the increased bias towards low SNR objects for the SNR2.5 subset.}
    \label{fig:rankedsnr}
\end{figure}

For each subset, we show the distribution of the distance scores in Figure \ref{fig:histscores}. The distance scores are used as the outlier scores of the objects, and the 100 weirdest objects in each distribution is indicated by the red line. With the increasing amount of objects going from left to right, we see that the shape of the distribution changes drastically. Still, all distributions show a tail of objects with a high outlier score that we will inspect. We also inspect the relation between the outlier score and the signal to noise ratio of the spectra. We already hinted on the correlation earlier in this section, which can clearly be seen in the \texttt{SNR2.5} subset results in Figure \ref{fig:rankedsnr}. Due to the bias towards lower signal to noise ratio spectra, finding the reason why those spectra have high outlier score becomes increasingly more difficult as they contain many fluctuations on a continuum level. \\

\begin{figure}[ht!]
    \centering
    \includegraphics[width=.95\linewidth]{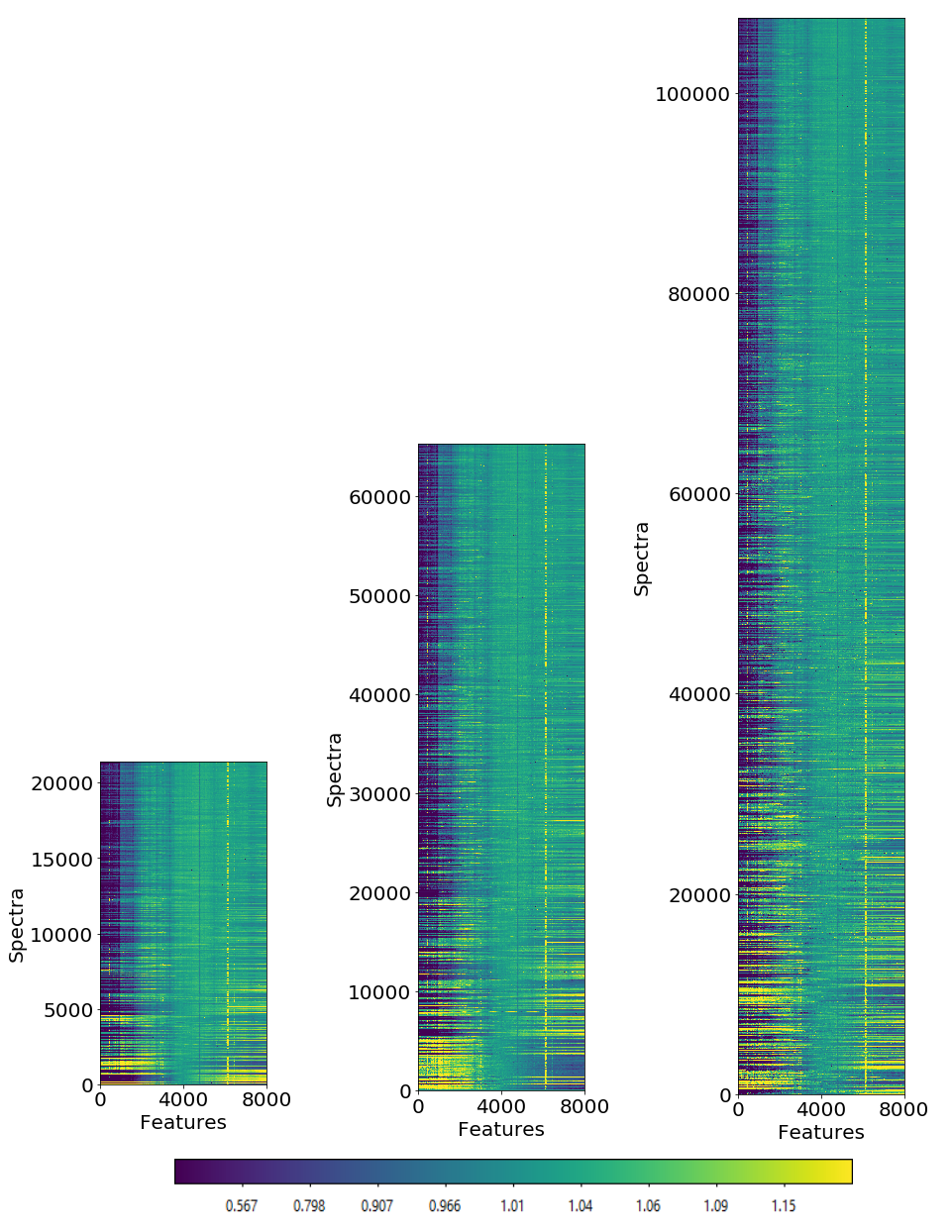}
    \caption{Normalized flux of the spectra of subsets (left to right) \texttt{SNR10}, \texttt{SNR5} and \texttt{SNR2.5} sorted on their outlier score in each subset. The weirdest spectra can be found at the bottom of each plot. Note the presence of clustering on extreme blue and extreme red spectra in all subsets. Due to the big amount of spectra in the subsets, our 250 inspected outliers are only a single line of pixels in these plots.}
    \label{fig:sortedall}
\end{figure}

In Figure \ref{fig:sortedall}, we show a nice visual representation of the spectra sorted on their weirdness scores after being run through the algorithm. In each color map, the objects with the highest outlier scores are found at the bottom of the map. This can also be seen by the more chaotic spectra found there. We also observe clustering on the continuum shape of extreme blue and red galaxies, as can be seen in Figure \ref{fig:sortedall}. Even with extensive testing and hyperparameter optimization, this problem persisted and this will be discussed in Section \ref{sec:discussion}. However, we could still find interesting outliers which are inspected in-depth in Section \ref{sec:outlier}.\\

\clearpage

\clearpage

\clearpage

\subsection{Variational Autoencoder}\label{sec:neuralnetwork}

Additional to the URF algorithm, we also experiment with the application of a reconstruction-based outlier detection technique on the spectra. This technique uses a \textit{variational autoencoder} \citep{Kingma_2013}, which can reduce data to a (very) low dimension and fully reconstruct the input. The difference between the reconstructed and input data can be used to determine an outlier score. The method was demonstrated on X-ray spectra in \citet{Ichinohe_2019} and was reported as a good approach to find outliers. On top of outlier detection, the variational autoencoder can also be used to generate synthetic spectra using the low dimensional representation in the trained network. Varying the input on the low dimensional space traces the important features in the data and can be used to construct spectra. The variational autoencoder can thus be used for outlier research, while it is also very useful for dimensionality reduction and feature importance analysis.\\

\begin{figure}[ht!]
    \centering
    \includegraphics[width=0.45\linewidth]{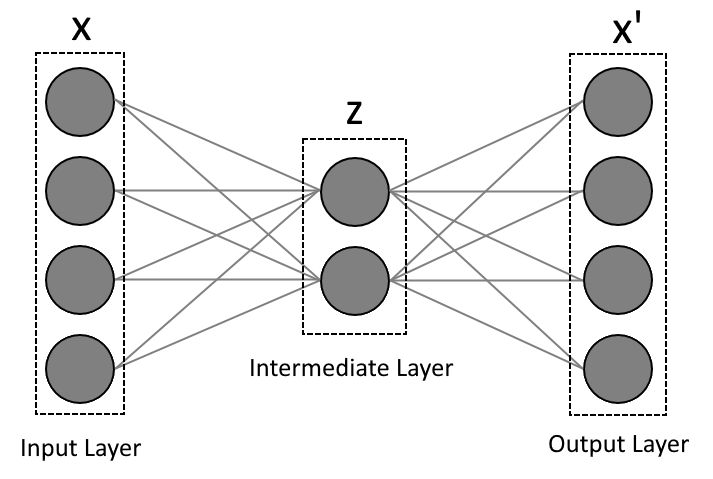}
    \caption{Basic autoencoder. The input layer \textbf{x} with dimension 4 is encoded to the intermediate layer \textbf{z} of dimension 2 via the connected neurons. The output layer \textbf{x'} is reconstructed from the lower dimension to the dimension of the input.}
    \label{fig:ae}
\end{figure}

The variational autoencoder is a variant of the normal autoencoder \citep{Hinton_2006}. An autoencoder is an unsupervised method and it utilizes a neural network architecture to learn a (non-linear) encoding of data. It consists of two connected parts: an encoder that maps the input data to a low dimensional representation and a decoder that can fully reconstruct the original data based on this representation. As the encoded dimension is (much) lower than the original dimension, the autoencoder has to learn the important features from the data to be able to fully reconstruct the input. The simplest form of an autoencoder is a neural network of three fully connected layers as shown in Figure \ref{fig:ae}. There is an input layer for the data, a single intermediate layer with a lower dimension than the input layer, and an output layer with the same dimension of the input layer. The layers are fully connected, as in that all the neurons in each layer are connected to all other neurons in the next layer via a (non-)linear function. The network is trained by adjusting the biases and weights of these functions to minimize a \textit{loss} function, which is often the reconstruction error between the input and the output. The dimensions of the layers, the specific function that is used between the layers and its biases and weights describe the full autoencoder. Two nice examples of the application of autoencoders in Astronomy are inspecting properties of stellar spectra \citep{Yang_2015} and classifying light curves of variable stars \citep{Tsang_2019}. These networks are composed of multiple layers between the input and the low dimensional layer to learn the more complex structures in the data, giving a more complex neural network than shown in Figure \ref{fig:ae}.

\clearpage

\begin{figure}[ht!]
    \centering
    \includegraphics[width=0.7\linewidth]{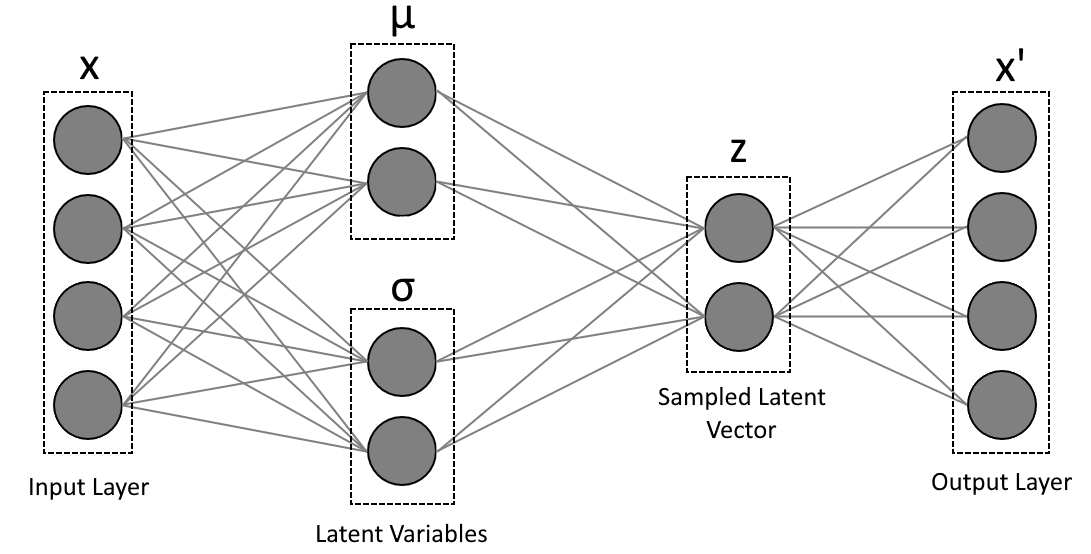}
    \caption{Variational autoencoder. The intermediate layers are two parallel layers representing the mean $\mu$ and log variance $\sigma$ of a Gaussian generative inference model. The decoder samples the latent vector \textbf{z} to reconstruct the input.}
    \label{fig:vae}
\end{figure}

Variational autoencoders are a modified version of autoencoders in two technical ways: At first, the encoder and decoder are not fully connected. Instead of encoding the input values to a lower dimension, the input gets mapped onto a distribution. The decoder samples from this distribution to reconstruct the input. This makes it that the neural network architecture has a Bayesian modeling approach, where the best model is learned for the best representation of the data. The second modification is that \textit{loss} of the network is not solely based on the reconstruction error, but also includes a loss term that restrains the mapped distribution of the low dimension latent variables. The addition of the mapped distribution gives that the layout of the variational autoencoder differs from the normal autoencoder, as can be seen in Figure \ref{fig:vae}. \\

The variational autoencoder consists of two coupled independent models: a recognition model as an encoder and a generative model as a decoder \citep{Kingma_2019}. The models work together to compress the input into a lower latent dimension. The latent dimension consists of latent variables, which are unobserved variables that represent the observed data. For example, a physical unobserved variable in the data could be the star formation phase of a galaxy, by learning from the ratios of the different emission lines. To relate the observed data with the unobserved latent variables, the decoder tries to model the underlying processes in the data via the model $p_\theta(\bm{x})$. This is done via the marginal likelihood
\begin{equation}
    p_\theta(\bm{x}) = \int p_\theta(\bm{x}, \bm{z})\ \text{d}\bm{z}\ ,
\end{equation}
where $p_\theta(\bm{x}, \bm{z})$ describes the decoder in the variational autoencoder and $\bm{z}$ denote the latent variables. The structure of the decoder is described as
\begin{equation}
    p_\theta(\bm{x}, \bm{z}) = p_\theta(\bm{z}) p_\theta(\bm{x} | \bm{z})\ ,
\end{equation}
with $p_\theta(\bm{z})$ the prior distribution of the latent variables. For the prior we use a Gaussian latent space. This is a simple model, yet works very good in most applications. An overview of deeper generative models is provided in \citet{Kingma_2019}.

\clearpage

\begin{figure}[ht!]
    \centering
    \includegraphics[width=0.7\linewidth]{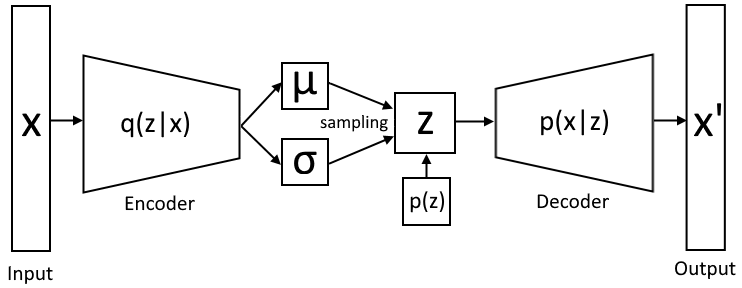}
    \caption{Bayesian representation of the variational autoencoder}
    \label{fig:bayesian}
\end{figure}

The encoder is an inference model $q_\phi(\bm{z}|\bm{x})$ that compresses the input data into two variables $\bm{\mu}$ and $\bm{\sigma}$, which represent the mean and log variance for the Gaussian inference model in the decoder. The variational autoencoder is trained by minimizing the following loss function, the \textit{evidence lower bound} (ELBO):
\begin{equation}\label{eq:elbo1}
    \mathcal{L}_{\text{vae}}(\phi, \theta, \bm{x}) = \mathcal{L}(\bm{x}, \bm{x'}) + D_\text{KL}(q_\phi(\bm{z}|\bm{x}) || p(\bm{z}))\ .
\end{equation}
This first part of loss function is the normal reconstruction loss of the autoencoder. The second part is the Kullback–Leibler (KL, \citealp{Kullback_1951}) divergence of the variational autoencoder, which traces two distances. By definition it traces the divergence between the approximate posterior and the true posterior of the data, representing how good the model can approximate the data. Additionally, it traces the difference between the encoder and decoder by comparing the encoder $q_\phi(\bm{z}|\bm{x})$ and the expected input for the decoder $p_\theta(\bm{z})$. A full Bayesian representation of the variational autoencoder is shown in Figure \ref{fig:bayesian}.\\

The decoder is a probabilistic inference model by sampling the latent vector $\bm{z}$ from the encoded means $\bm{\mu}$ and log variance $\bm{\sigma}$. The sampling gives that the reconstructed data can be made from a continuous distribution, resulting in the ability to interpolate between the learned features. For example, if a variational autoencoder is trained on galactic spectra, then the latent dimension can be used to generate synthetic spectra of any shape by choosing specific values as input. In the ideal case, the latent variables would represent real physical parameters of galaxies that can be tuned, but that requires a very complex network. Just as with the normal autoencoder, if the variational autoencoder is applied to high dimensional data, then there are multiple layers in the encoder and decoder to learn the complex patterns in the data.\\

We use a modified version of the variational autoencoder called Information Maximizing Variational Autoencoders (InfoVAE, \citealp{Zhao_2017}). This method was used as an initial demonstration of variational autoencoders on galaxy spectra in \citet{Portillo_2020} and showed good results in greatly reducing the dimension of the spectra to only 6 variables and finding a few outliers. InfoVAE addresses two potential problems with the application of variational autoencoders on big data problems. At first, the KL divergence is not strong enough to be able to map the different kinds of spectra to a representative distribution. This is especially a problem if the input dimension is much larger than the latent space dimension, as in our case. Also, the KL divergence term does not take into account that for a complex network, the encoder can always match the prior of the decoder. This means that the network did not learn any meaningful latent variables that describe the features of the data and will result in similar findings as a normal autoencoder. 

\clearpage

The modifications of InfoVAE are applied via adjustments of the ELBO function. An extra term \textit{Maximum Mean Discrepancy} (MMD) is added with additional weighting factors. This term is based on the idea that two distributions are identical if their moments are the same. It compares the outcome of the encoder $q_\phi(\bm{z})$ with the expected distribution $p_\theta(\bm{z})$ and forces them to be similar via training of the variational autoencoder. The exact definitions of $D_\text{KL}$ and $D_\text{MMD}$ as applied in our algorithm are thoroughly described in \citet{Kingma_2019} and \citet{Zhao_2017} and briefly shown in Appendix B. The full ELBO function to minimize is given by

\begin{equation}\label{eq:elbo}
    \begin{aligned}
    \mathcal{L}_{\text{InfoVAE}}(\phi, \theta, \bm{x}, \alpha, \lambda) = 
    &\ \mathcal{L}(\bm{x}, \bm{x'})
    + (1-\alpha)D_\text{KL}(q_\phi(\bm{z}|\bm{x}) || p_\theta(\bm{z})) \\[0.1em]
    &\quad+ (\alpha + \lambda - 1)D_\text{MMD}(q_\phi(\bm{z})||p_\theta(\bm{z})) \ ,
    \end{aligned}
\end{equation}\\

where the exact value of the weighting values $\alpha$ and $\lambda$ have to be found via hyperparameter optimization. As recommended in \citet{Zhao_2017} we will use $\alpha = 0$, which gives that the KL divergence is the same as in \eqref{eq:elbo1}. For $\lambda$ we test different values in the range $\lambda = 5-15$ to search for the best latent variable representation.

\subsubsection{Training}
We train the InfoVAE with the \texttt{SNR5} subset that was also used with the URF implementation. This subset contains around 63839 good spectra with high signal to noise ratio, giving a big data set with high quality data for this experimental method. We test two different networks, one with the input dimension of 8000 points as used with the URF, and one with the input dimension of 1000 points as used in \citet{Portillo_2020}. Lowering the input size reduces the information put into the network, but greatly improves the training speed. We compare the reconstruction capability of the networks and look if they can used for outlier detection. \\

The neural network structure in the variational autoencoder is very important for the performance of the models and has to be carefully tuned to get the best results. We tested different networks with multiple layers of varying sizes. The lowest reconstruction errors were found with three fully connected layers with the dimensions 1000-500-100 converting the 8000 input points to 6 latent variables. Even with the low number of 6 latent variables galactic spectra can be fully reconstructed with low errors as shown in \citet{Portillo_2020}. The network parameters used in their work are very similar to our optimal network, suggesting that we also have a good network. For the network with only 1000 input points the network structure is the same, but using the first encoding layer as the input layer omitting the first encoding step.\\ 

We build the neural network with Python using Keras\footnote{\url{https://www.tensorflow.org/api_docs/python/tf/keras}}, a high-level API for the Tensorflow 2 platform. An overview of the full network is shown in Figure \ref{fig:neuralnet} in Appendix C. The layers are connected with the non-linear ReLu activation function \citep{Relu_2010} that can learn the non-linear relationships between the input points. The output layer uses a linear activation as some spectra can have negative values due to normalization. The network is trained by minimizing the loss function shown in \eqref{eq:elbo}, using the built-in Adam optimizer with default values. The code can be found on the Github page.

\clearpage

Before training, we split the subset into a training set and validation set. We train on 95\% of the spectra and use the remaining 5\% validation data to probe the performance of the network on unknown data. We stop training if the performance on the validation set does not increase for more than 30 epochs. For both input sizes, this happens at around 400 epochs given a batch size of 2048. \\

\begin{figure}[ht!]
    \centering
    \includegraphics[width=\linewidth]{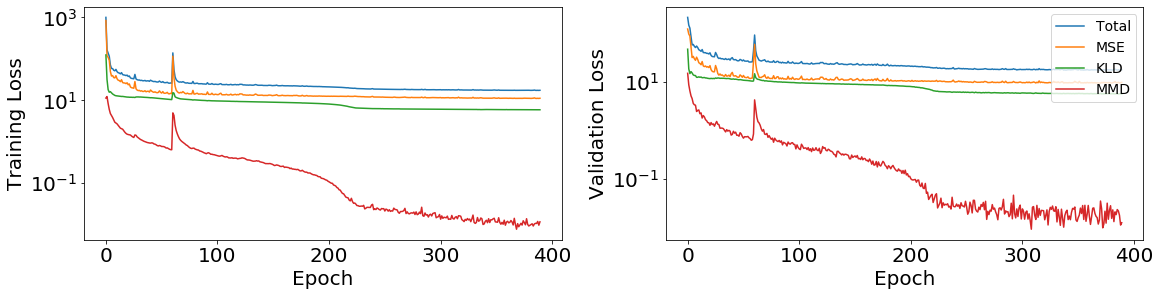}
    \caption{Loss minimization during training for both the training and validation data}
    \label{fig:loss}
\end{figure}

The progress of loss minimization during the training is shown in Figure \ref{fig:loss}. The loss is composed of three components, the standard MSE loss, the KL Divergence loss, and the MMD loss. According to the theory, the MMD loss should be higher than the KLD loss as it is supposed to support the convergence. However, this is not the case, making the MMD loss obsolete in our training. This difference between theory and our observed MMD loss is due to different scaling parameters in the loss function. Unfortunately, if we apply those \textit{correct} scaling parameters, then the MMD loss is dominating the loss function by a few magnitudes prohibiting the neural network to learn anything from the data. The best performance was found without using the MMD loss during training. This is different as reported in \citet{Portillo_2020}, but as they did not show the loss minimization plots during training, we can not make a direct comparison between the performances.

\subsubsection{Latent Space Representation} 
\begin{figure}[ht!]
    \centering
    \includegraphics[width=\linewidth]{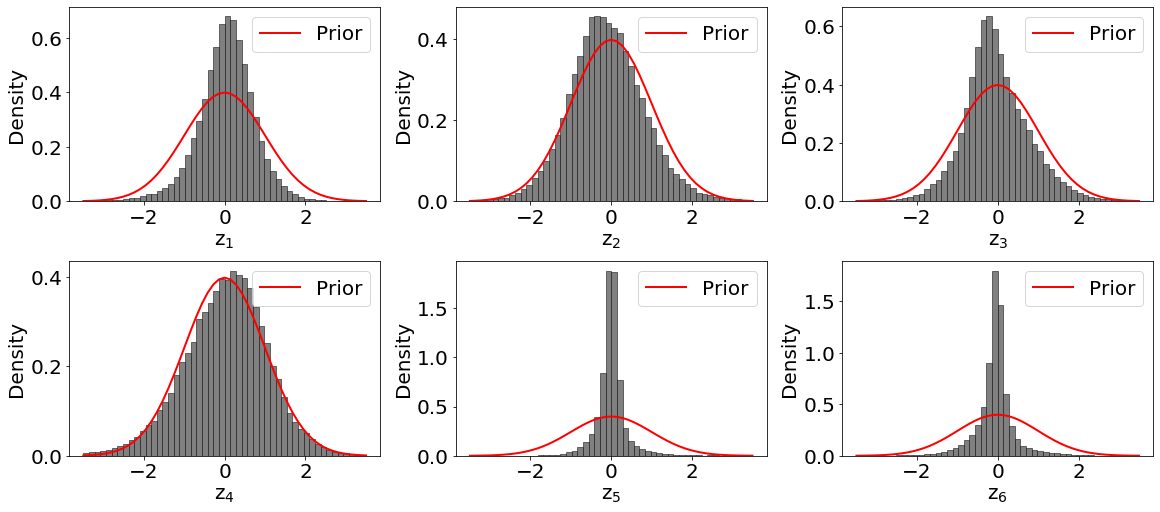}
    \caption{Distribution of the values of each latent variable. If the distribution is similar to the prior of the Gaussian inference model, then the latent variable traces is used to trace important features.}
    \label{fig:latenthist}
\end{figure}

The trained variational autoencoder consists of two models, the recognition model and the inference model. We can split these models and use one at the time to either convert the spectra to their latent space representation using the encoder, or generate synthetic spectra from (random) latent variables using the decoder. We first inspect the encoded latent space representation of the data set to see if the network has learned features in the data. All the spectra are converted to their 6-dimensional latent space vector, where ideally each variable or a combination of variables trace the features in the data. We compute the mean and variance of each individual latent variable and inspect the distribution of the values. In Figure \ref{fig:latenthist} we show the distribution of latent space values for all spectra. If the distribution of a latent variable is similar to the normal distribution $\mathcal{N}(0, \textit{I})$ of the prior, then the latent variable represents important features of the data set. If it is not similar, it is either not used for reconstruction, or used as a support variable for a complex relationship in the data. For 4 of the 6 latent variables, we see a good or similar shape for the distribution of the latent variable indicating learned features. 

\clearpage

\begin{figure}[ht!]
    \centering
    \includegraphics[width=\linewidth]{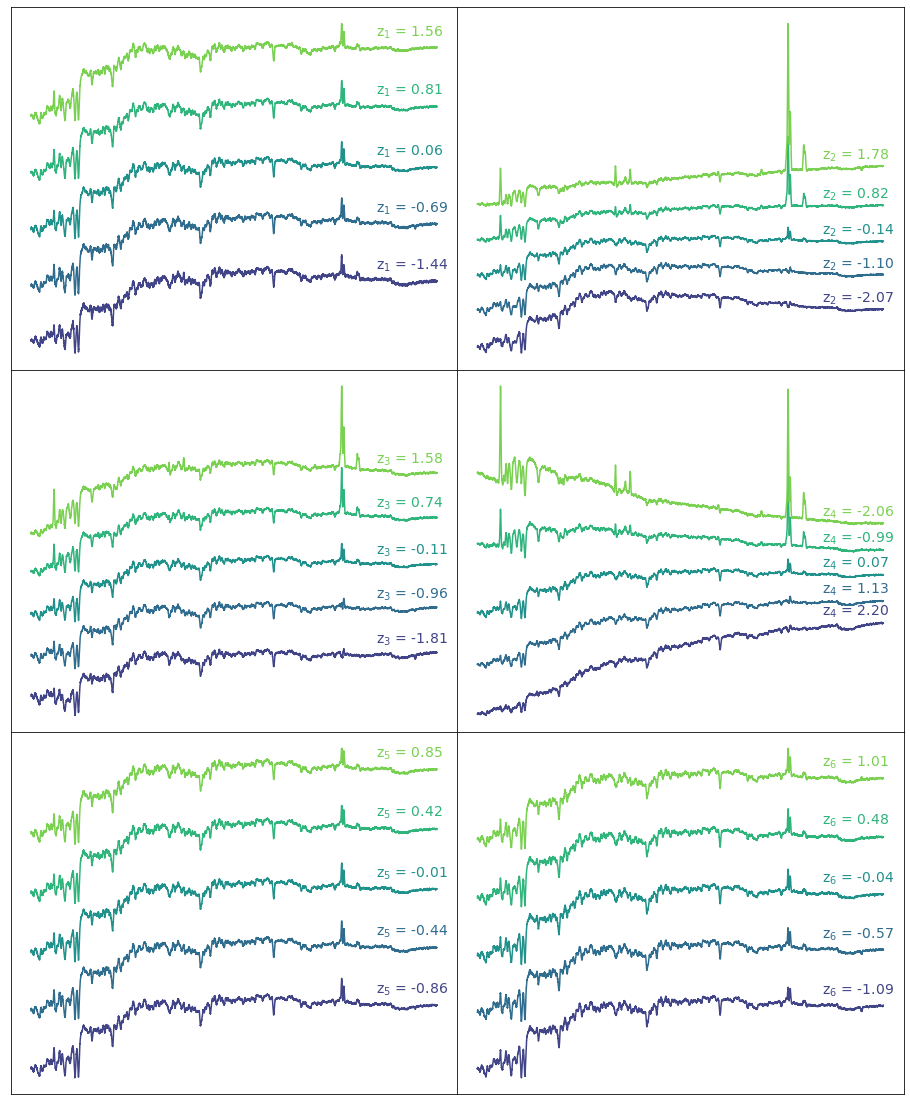}
    \caption{Synthetic spectra generated from the trained variational autoencoder. In each plot, we only change a single latent variable at a time. This shows the relation between the inference model and the latent space. Note that is a similar representation as in Figure 14 in \citet{Portillo_2020} to compare the performances of the networks.}
    \label{fig:syntheticvar}
\end{figure}

We generate synthetic data using the information from the distributions and apply it on the decoder. If we take the mean of the values for each latent variable and decode it to a spectrum, the decoded spectrum should represent the most common type of spectrum found in the data. By tuning a single latent variable at a time using the variance of the distribution, we can see the features that each individual latent variable traces. In Figure \ref{fig:syntheticvar} we show generated spectra from the latent space by changing only a single variable at a time and using the mean of the other latent variable values. The value of the varied latent variable is shown for the individual plots. This method of displaying the latent variable changes is directly taken from \citet{Portillo_2020}, and we can compare the different features found by our network and theirs. \\

For 3 of our latent variables ($z_2$, $z_3$ and $z_4$), we can easily see what features of the spectra depend on the values of the latent variables as line intensities or continuum shapes vary a lot. The other 3 do not show major differences at first sight, but show more subtle differences in absolute values on a continuum level. The difference between the synthetic spectra of latent space variable $z_1$ is difficult to see in the representation of Figure \ref{fig:syntheticvar}, but the continuum is either slightly flattened or curved depending on the value. For $z_5$ and $z_6$ we see a slight difference in the ratios between different emission lines and different heights of the continuum flux.\\

With both Figure \ref{fig:latenthist} and Figure \ref{fig:syntheticvar} we can see that the latent variables that have their distribution matching the prior show the most obvious features when they are varied. For example, latent variable $z_4$ matches the prior perfectly and shows significant variation in the emission line strengths and continuum shape. Unfortunately, we could not train the network to have all 6 latent variables matching the prior. As shown in Figure \ref{fig:loss} and discussed in 3.2.1, the MMD loss should have helped to force the distributions to match the prior, but this only resulted in worse performance when applied in our algorithm. In \citet{Portillo_2020}, all latent variables trace more obvious features such as line broadening or very big changes in line emission ratios. Unfortunately, even with extensive comparison between our method and theirs we could not reproduce this.

\clearpage

\clearpage

\begin{figure}[ht!]
    \centering
    \includegraphics[width=0.8\linewidth]{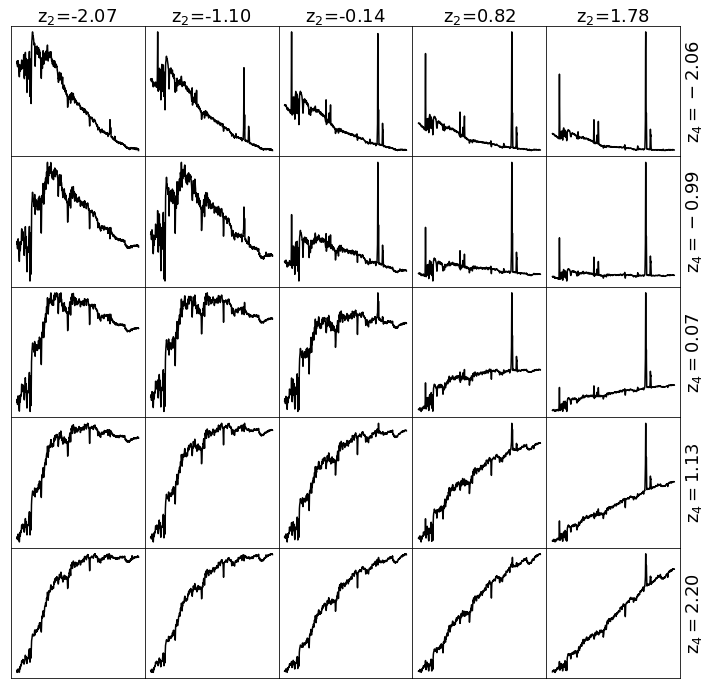}
    \caption{Synthetic spectra generate from the trained variatonal autoencoder changing multiple latent variables. With only two latent variables a huge range of spectra can be generated. Note that no y-axis values are given as the absolute flux values are not real. We only want to show the different shapes and features of the spectra using these two latent variables.}
    \label{fig:syntheticcross}
\end{figure}

In Figure \ref{fig:syntheticcross} we cross-correlate two of the latent variables ($z_2$ and $z_4$). As shown, with only these two latent variables we can already represent a good amount of types of spectra ranging from blue galaxies to red galaxies, including and excluding emission lines. All 6 latent variables trace some or multiple common features of the spectra, and combining them lets us generate most types of galaxies found in our GAMA subset. As outliers have either uncommon features or weird lines, we are confident that the variational autoencoder can not reconstruct those correctly resulting in high outlier scores.

\vspace{1cm}

\clearpage

\subsubsection{Reconstruction Capability}\label{sec:recon}
We have shown that the variational autoencoder is capable of generating synthetic data given only 6 parameters. But it is also important to check if the spectra are reconstructed correctly. This will also be used for outlier detection, as uncommon features are not learned by the neural network. We will look at the reconstruction of a few selected spectra, including either very common, or uncommon shapes. We also compare the reconstruction capability between the network with 8000 and 1000 input points. \\

\begin{figure}[ht!]
    \centering
    \includegraphics[width=\linewidth]{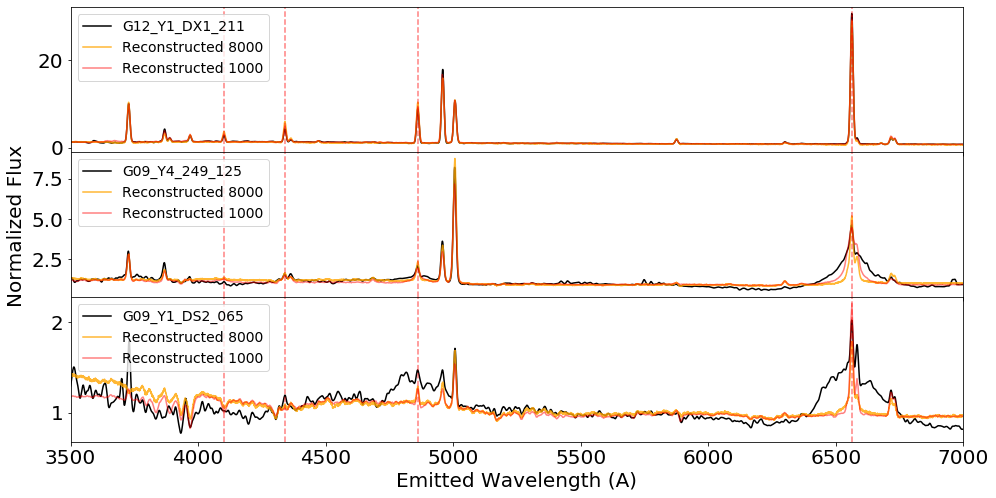}
    \caption{Reconstruction capability of the trained variational autoencoder for increasingly more complex Balmer lines. }
    \label{fig:balmerreconstruction}
\end{figure}

First, we look at three spectra with clear emission lines visible to see if those are reconstructed correctly. In Figure \ref{fig:balmerreconstruction} the spectra are shown including the reconstructed interpretation using the variational autoencoders with the inputs of 8000 data points or 1000 data points. From top to bottom we show increasingly more complex emission lines. Overall, the continuum is traced very well for each spectrum. Spectrum \texttt{G12\_Y1\_DX1\_211} has high emission for the most common star formation lines and shows good results in the reconstructed version. In \texttt{G09\_Y4\_249\_125} the Balmer emission lines are broadened due to activity in the galaxy. The reconstructed version can trace some the broadening of the lines, but does not reconstruct it completely correct. For \texttt{G09\_Y1\_DS2\_065}, the additional complex structure at the Balmer lines is not reconstructed at all, giving a high reconstruction error. The more the complexity in the emission lines, the more the reconstruction error tracing a possible outlier. \\

\begin{figure}[ht!]
    \centering
    \includegraphics[width=\linewidth]{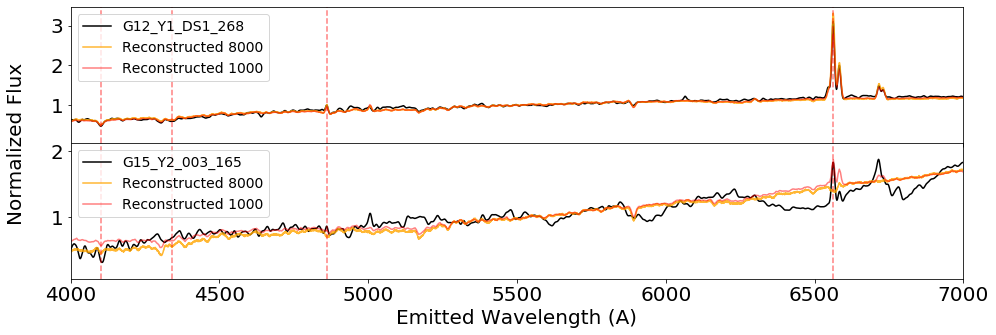}
    \caption{Reconstruction capability of the trained network for a blended object. The top spectrum is a very good reconstruction of a single galaxy, while the bottom spectrum is a galaxy with an M-dwarf star in the foreground. While the x-axes indicated the rest-frame wavelength, the foreground flux of the M-dwarf is redshifted giving the redshift of the background galaxy. }
    \label{fig:blendreconstruction}
\end{figure}

In Figure \ref{fig:blendreconstruction} we show the spectra of a very 'normal' galaxy and a blended system in which an M-dwarf sits in front of a galaxy. Spectrum \texttt{G12\_Y1\_DS1\_268} is reconstructed very accurately, where only very small fluctuations on the continuum level are missed. On the other hand, spectrum \texttt{G15\_Y2\_003\_165} shows a big difference between the input and the reconstruction. This is due to the foreground spectrum of the M-dwarf interfering with the flux of the galaxy. For all blends, as the full spectrum is redshifted using the emission lines of the background galaxy, the features of the M-dwarf will always be in seemingly random locations.

\vspace{0.5cm}

These examples of uncommon spectra show that the trained variational autoencoder can reconstruct common features in the data very accurate, but fails to reconstruct extra complex structures or blended objects. The reconstruction errors of these objects are higher than those for normal spectra, giving us great confidence in the usage of the variational autoencoder in the search for outliers. Looking at the reconstructed spectra from variational autoencoder with 8000 input points and 1000 input points, we do not see any major differences in reconstruction errors. On closer inspection, we do see an increased performance of the 8000 input network at the heights of the emission lines. We can show many more examples of bad reconstructed spectra, but we will refer the reader towards Section \ref{sec:outlier}, in which the types of outliers are discussed.

\subsubsection{Outlier Scores}\label{sec:outliersmet}
The variational autoencoder is used as a reconstruction based outlier detection method. This uses the fact that it learned the important features of the majority of the data, while lacking the knowledge of weird shapes or features found in outliers. Besides computing the reconstruction error, we also look at the latent space representation of all spectra. This low dimensional latent space contains, just as with the distance matrix of the URF method, important information of the spectra. We can use this information to find outliers inside the latent space, by looking at points far from the mean of the distribution. \\

We start by looking at the reconstruction error of the spectra defined by a function that computes the differences between the input flux values and the reconstructed flux values. Many existing metrics can be used, but one could also make up one on their own. We use a comparable metric as used in \citet{Ichinohe_2019}, where they defined the outlier score via a statistic logarithmic chi-square metric. 

\clearpage 
\begin{figure}[ht!]
\begin{subfigure}{0.45\linewidth}
    \centering
    \includegraphics[width=\linewidth]{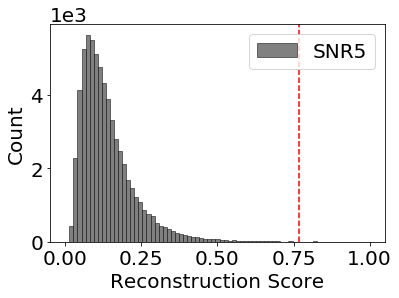}
    \caption{Outlier scores computed via the reconstruction error of Equation \eqref{eq:recon}}
    \label{fig:recondist}
\end{subfigure}\hfill
\begin{subfigure}{0.45\linewidth}
    \centering
    \includegraphics[width=\linewidth]{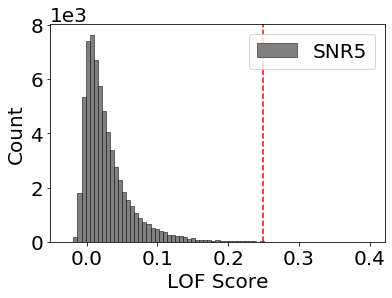}
    \caption{Outlier scores computed from the latent variables using LocalOutlierFactor method.}
    \label{fig:lofdist}
\end{subfigure}
    \caption{Outlier score distributions based on the output of the variational autoencoder. The 100 weirdest objects are found on the right side of the red lines.}
    \label{fig:nnscores}
\end{figure}
\vfill

\begin{figure}[ht!]
    \centering
    \includegraphics[width=\linewidth]{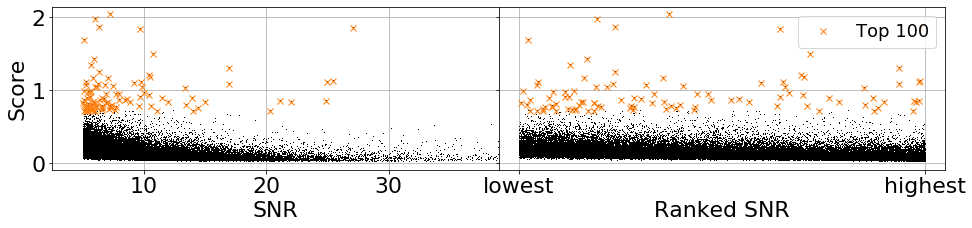}
    \caption{Relation between the reconstruction outlier score and the signal to noise ratio of the spectra. In comparison to the URF, we do not see any bias towards low signal to noise ratio spectra.}
    \label{fig:reconsnr}
\end{figure}

We use a slightly modified version of their metric, omitting the logarithmic scaling, and apply 

\begin{equation}\label{eq:recon}
    \text{median}\left\{ \left(\frac{(I - R)}{\sqrt{I}}\right)^2 \right\}  \ ,
\end{equation}

on the reconstructed spectra of the \texttt{SNR5} subset as the outlier score, where \textit{I} is the input and \textit{R} the reconstructed output. Another suggested metric in \citet{Ichinohe_2019} is the maximum difference between the reconstructed and input via $\max(|R-I|)$, which can trace a single high emission line that are not correctly reconstructed. Application of this simpler metric on our data showed outliers that consist almost exclusively of spectra with a single extremely high emission line most probable originating from bad reduction or a cosmic ray event. We normalize the output of \eqref{eq:recon} such that weird spectra are assigned a score close to 1. The score distribution is shown in Figure \ref{fig:recondist}, which shows a high peak of spectra with a similar reconstruction error and a tail of outlying spectra with a high reconstruction error, indicating the outliers. Just as with the URF, we also look at the relationship between the outlier scores and the signal to noise ratio of the spectra to investigate any bias. In Figure \ref{fig:reconsnr} we can see that there is no bias towards lower signal to noise ratio spectra as was found with the URF method.\\

\vfill

\clearpage

We also look at the latent space variables that trace the encoded features of the input spectra. As the variational autoencoder is not trained to encode uncommon or weird features correctly, spectra containing those can end up with very different latent space vectors in comparison to the normal spectra. Using the latent space variables to find outliers is computational beneficial as not all the spectra have to be reconstructed. However, the latent space has to be explored with a simple unsupervised clustering method instead, adding up to the computational time. A second disadvantage is that we can not see why a spectrum is assigned a high outlier score, so reconstruction is still necessary in the end. We still want to explore this method to examine the capabilities of outlier detection of spectra in an encoded space.\\

\begin{figure}[ht!]
    \centering
    \includegraphics[width=\linewidth]{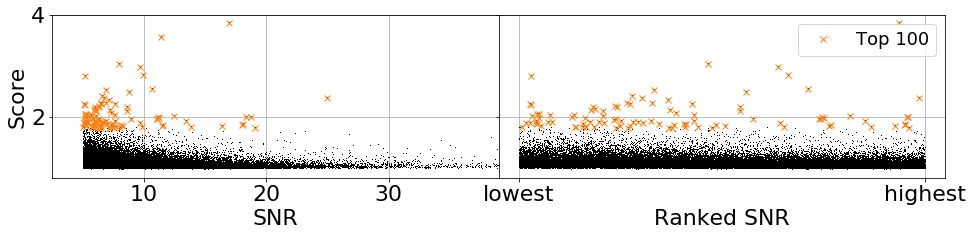}
    \caption{Relation between the LOF score and the signal to noise ratio of the spectra. In comparison to the URF, we do not see any bias towards low signal to noise ratio spectra.}
    \label{fig:lofsnr}
\end{figure}

In \citet{Portillo_2020}, they briefly suggested and used the unsupervised outlier detection method LocalOutlierFactor (LOF, \citealp{Breunig_2000}), an easy to use algorithm found in the scikit-learn package in Python\footnote{\url{https://scikit-learn.org/stable/modules/generated/sklearn.neighbors.LocalOutlierFactor}}. In short, the method computes the local density of objects by looking at its \textit{k} nearest neighbors. For each object, its local density is compared to that of its neighbors and a score is assigned based on this difference. The Python based method outputs a \textit{negative outlier factor} for each object. We define our outlier score with $\log[-1\cdot\text{NOF}]$, where NOF is the negative outlier factor, and get the score distribution as shown in Figure \ref{fig:lofdist}. Normal objects can be found at an outlier score of 0, while outlying objects have higher values. Similar to the reconstruction score, a high peak with normal galaxies is found with a tail of outlying galaxies. The relation between the LOF score and signal to noise ratio of each spectrum is shown in Figure \ref{fig:lofsnr}, where again no correlation is found between the outliers and its quality. \\

\begin{figure}[ht!]
    \centering
    \includegraphics[width=0.7\linewidth]{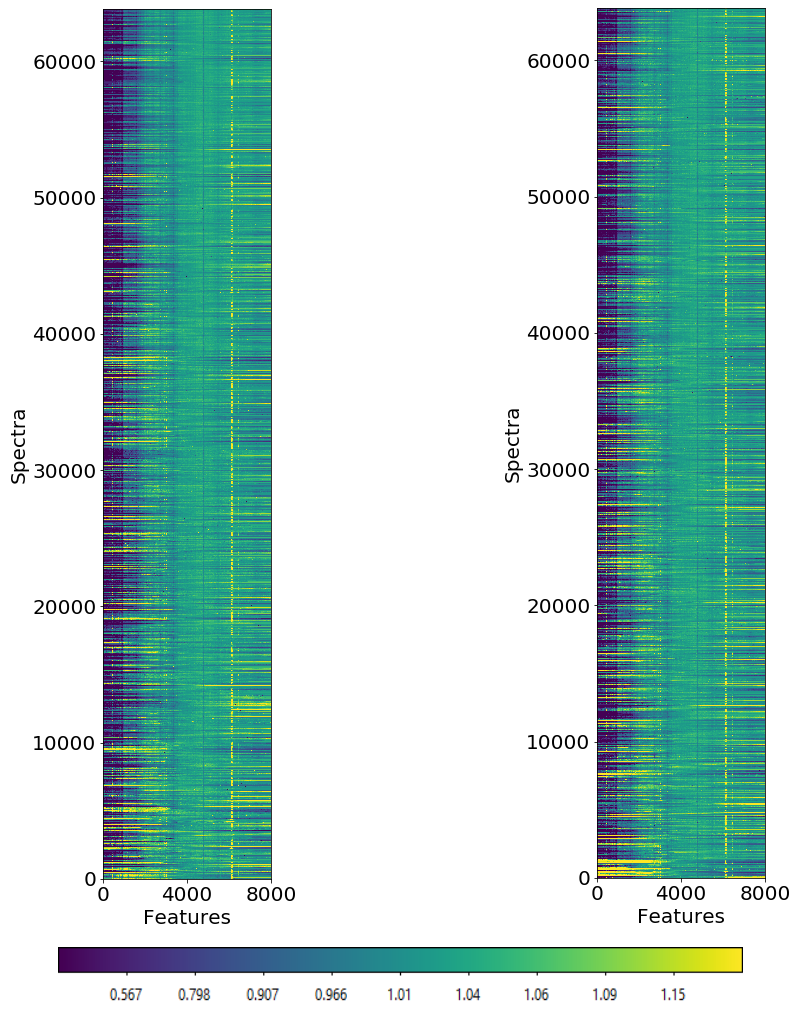}
    \caption{Normalized flux of the spectra of subset \texttt{SNR5} sorted on their outlier score. The weirdest spectra with the highest outlier scores are found at the bottom of the maps. \textit{Left:} sorted on outlier score determined via reconstruction error. \textit{Right:} sorted on outlier score determined via LocalOutlierFactor.}
    \label{fig:sortedallnn}
\end{figure}

We show the spectra sorted on their outlier score for both the reconstruction based method and the latent space method in Figure \ref{fig:sortedallnn}. In the overview, we see no obvious clustering on similar types of spectra as was found in the overview in Figure \ref{fig:sortedall}. The main reason for this difference is that the URF outlier algorithm method is distance-based, using pair-wise distances of the spectra as the basis for the outlier scores. The variational autoencoder reconstructs all spectra individually lacking any pair-wise information, resulting in less clustering of similar spectra.\\

\clearpage

\vspace{1cm}

\begin{figure}[ht!]
    \centering
    \includegraphics[width=0.6\linewidth]{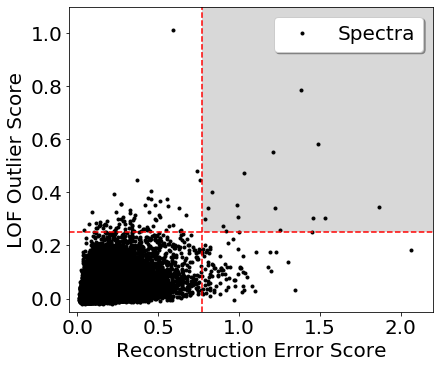}
    \caption{Correlation between outlier score methods for the variational autoencoder}
    \label{fig:scorecor}
\end{figure}

At last, we show the correlation between the outlier scores computed from the latent space vectors and the outlier scores computed from the reconstruction errors in Figure \ref{fig:scorecor}. While both method find their own types of outliers, we do see an overlap of outliers in the grey colored area. We also see that there is not a very strong correlation between the scores. This suggests that, while the latent space vector represents the important features of the spectra, the decoder adds another dependency to the types of outliers we find. Also, while the encoder will map known features to a specific latent space vector, uncommon or weird features might be mapped to a random value as it is unknown to the encoder. Another possible reason for the difference is the specific metric we used to compute the outlier scores. We inspect the types of outliers found by both methods in Section \ref{sec:outlier}.\\

\clearpage
\clearpage
\section{Outliers}\label{sec:outlier}

We inspect the spectra that have the highest outlier scores. The primary search for outliers was done with the URF algorithm, so we inspect those first and compare them with the results of the variational autoencoder afterward. As we used multiple subsets, we have to define a method to find the weirdest spectra in the GAMA survey. First, we look at the highest quality subset \texttt{SNR10} and inspect the weirdest 100 spectra. Afterward, we inspect the \texttt{SNR5} subset for another 100 weird spectra we did not find in the \texttt{SNR10} subset. At last, we inspect the \texttt{SNR2.5} subset, in which we had a lot of difficulty in finding out why some spectra are assigned a high score. Therefore, we only inspect the weirdest 50 spectra from the lowest quality subset. This gives a total of 250 unique weird spectra when combined. We will discuss these decisions in Section \ref{sec:discussion}. We expect to find many outliers that can be explained, but hope to also find interesting outliers that we do not understand. There are definitely more interesting outliers in the GAMA survey that we do not inspect in this Section, but these are up for the reader to find as we provide a full list the outlier scores on the Github page.

\subsection{Unsupervised Random Forest}\label{sec:urff}

Before diving into the outliers, we note that a lot of the spectra are weird due to instrumental or reduction errors. These spectra naturally came up with the highest weirdness scores and had to be removed manually to ensure interesting outliers. This iterative procedure resulted in a total of 2926 spectra that showed fringing, not earlier reported by the GAMA team. Moreover, 1283 spectra showed a bad splice in the continuum and were unusable, also not earlier reported. Including the ND1 field and (sky) reduction errors, we conclude that at least 4.5\% of the AAOmega spectra in the GAMA survey can not be used for big data projects, unless a reliable method can be developed to correct all the spectra. This percentage is slightly higher as earlier reported in \citet{Hopkins_2013}. We attempted to defringe the fringed spectra, but this introduced additional artifacts in the data due to the differences between the fringes. The full list of bad spectra is provided on the Github page to be used in other projects. \\

It can be very difficult to determine why a spectrum was assigned a high outlier score, as it might not be obvious why the algorithm has set a specific score for a spectrum. We try our best at investigating what makes the outlying spectra interesting and comment on its presence in the literature using SIMBAD \citep{simbad}. First, we classify a great amount of spectra via visual inspection, as there are obvious features like extreme emission lines or very broad features. For the more difficult spectra, we look in-depth at the flux values of the known emission lines, additional emission lines, or other weird features in the continuum. We find many similar types of spectra as in \citet{Baron_2016} and the classification groups will be very similar. The different types of outliers are extensively described in the next parts and tables of the inspected spectra are provided in Appendix D. Overall, we divide the spectra into 4 groups that define the main characteristics of the spectra. The first group is composed of spectra with unusual velocity structures or broadened emission lines, tracing different active processes in the galaxies. The second group consists of spectra with extremely strong emission lines, uncommon emission lines, or unusual emission lines ratios. Some spectra show characteristics of two objects types, like a star in front of a galaxy, and are grouped as blends. The remaining spectra are grouped mainly on their unusual continuum shapes, or contain weird non-physical features.

\clearpage

\begin{figure}[ht!]
    \centering
    \includegraphics[width=\linewidth]{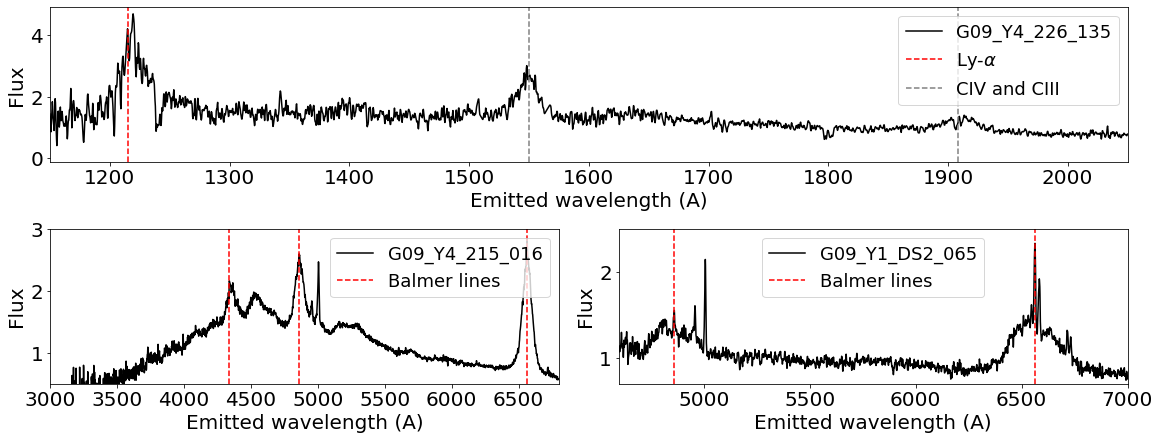}
    \caption{Three types of spectra with unusual features from kinematic processes inside the galaxy. \texttt{G09\_Y4\_226\_135} is a QSO at relatively high redshift showing broad Ly-$\alpha$ and Carbon emission. The spectra \texttt{G09\_Y4\_215\_016} and \texttt{G09\_Y1\_DS2\_065} show broadening of, or additional structure at the Balmer lines tracing active processes in the galaxy.}
    \label{fig:kinematics}
\end{figure}

\subsubsection{Unusual Velocity Structure}\label{sec:unusualvel}
The first and most obvious outliers are due to different kinematic processes in the galaxies giving rise to unusual velocity structures or line broadening. These are the easiest to inspect as there is clearly some additional structure in an otherwise normal spectrum. Among these spectra, we find many Quasi-Stellar objects (QSOs), different Active Galactic Nuclei (AGNs), and a few spectra that show minor broadening in their emission lines.\\

In total, we have 50 spectra with features that can be traced back to different kinematic processes going on in the galaxies. We find 14 Quasi-Stellar Objects at relative high redshift with Carbon and Silicon emission, whereof 7 objects are Lyman-$\alpha$ emitters. These high redshift spectra are actually not found due to the emission lines, but are assigned a high outlier score as they are a flat line in our interpolated wavelength range. Of the spectra in our wavelength range, we find 27 spectra with additional (complex) structure at either H$\alpha$ or all the Balmer lines. At last, 9 objects have slight broadening at the most common emission lines.\\

The 14 high redshift QSOs are easily found, as they fall outside of our interpolation range. Note that the term 'high redshift' is only relative to the other objects in the GAMA survey. These objects could also be found by looking at the redshift values, but the GAMA survey provides no information about the flux of the emission lines in these spectra and classifying still has to be done via visual inspection. These spectra show broad emission lines for Carbon, while the spectra at slightly higher redshift also show broad Silicon, Lyman-$\alpha$, and even Lyman-$\gamma$ emission. Out of these 14, only two are earlier reported in the 2dF QSO survey \citep{Croom_2004} and 9 are flagged as QSO in the GAMA database. We show an example of the QSO spectrum \texttt{G09\_Y4\_226\_135} in Figure \ref{fig:kinematics} and can easily see the broad emission lines.\\

\clearpage
Additional to high redshift QSOs, we find 27 spectra with unusual velocity structure features originating from active galaxies. Of these spectra, 19 have extreme broadening of the Balmer lines and 8 show a more complex structure around the Balmer lines. Two examples of broadening and an asymmetric additional structure at the Balmer lines are shown in Figure \ref{fig:kinematics}. The extreme broadening of Balmer lines is due to active galactic nuclei, of which most spectra are of Seyfert-1 galaxies. Additional to the broadened Balmer lines, we also observe other velocity structures in the continuum of \texttt{G09\_Y4\_215\_016}. These can not directly be traced to specific emission lines and probably trace more outflows of the active galaxy. The complex structure in \texttt{G09\_Y1\_DS2\_065} is found at both the H$\alpha$ and H$\beta$ lines, indicating a relation between the structure and the Balmer lines. A different spectrum with this structure was also found in \citet{Baron_2016} and was modeled to be a combination of both broad Balmer emission seen in Seyfert-1 galaxies, and Balmer absorption in supernova ejecta \citep{Faran_2014}. Most of the spectra are already reported in the quasar catalogs of \citet{Rakshit_2017} and \citet{Toba_2014}, but 9 of these spectra have either no notable references or are classified as stars. \\

A special mention is the spectrum \texttt{G09\_Y6\_090\_043}, which showed a broad emission structure at the MgII emission line not recognized by GAMA. This line is often very variable \citep{Homan_2019}, so this spectrum could be an example of an observation of an event that only exists on a short timescale. However, as there are no other spectra of this galaxy in other surveys, this variability can not be checked.\\

At last, we find 9 spectra with a slight broadening of emission lines or minor additional structure in the H$\alpha$ and NII structure. The spectrum \texttt{G12\_Y1\_DS2\_080} showed minor broadening of high emission lines, tracing star formation and some activity. This spectrum is of a rare type called \textit{Green Bean} Galaxies and is one of the 14 spectra studied in \citet{Prescott_2019}. These Green Beans have a very blue continuum, making them a subtype of Type 2 AGNs. 

\begin{figure}[ht!]
    \centering
    \includegraphics[width=\linewidth]{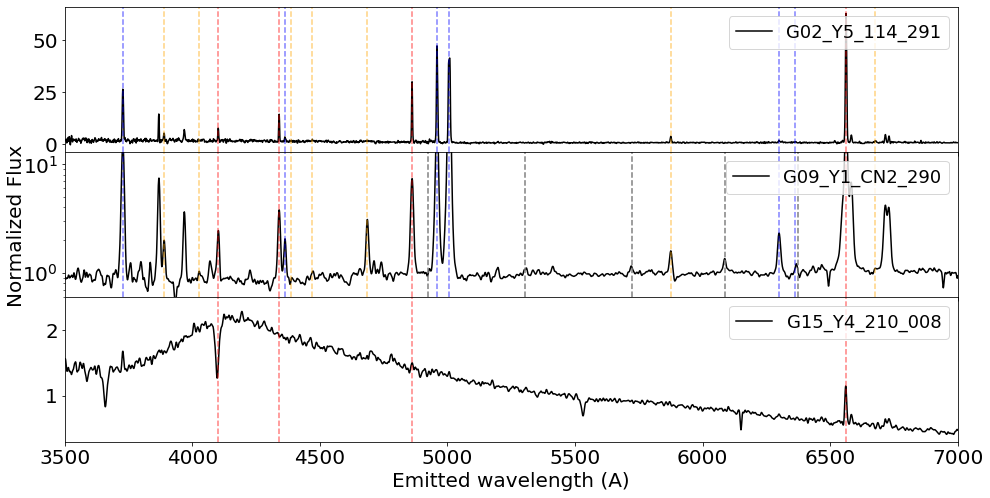}
    \caption{Three types of spectra classified based on their emission lines. From top to bottom: Extreme star formation and BPT outlier, Iron emission lines presence, and a post-starburst spectrum. Emission lines: red = Hydrogen, blue = Oxygen, orange = Helium, grey = Iron.}
    \label{fig:lines}
\end{figure}

\clearpage

\subsubsection{Emission Lines}\label{sec:emissoinlines}
Most of the outliers show a range of different emission lines with different intensities, but do not show any obvious unusual velocity structures. This big group is composed of spectra that have very strong or uncommon emission lines, including spectra with unusual emission line ratios tracing the different star formation phases. We find 114 spectra with very high emission lines of the common elements: Hydrogen, Helium, Oxygen, Nitrogen, and Sulfur. Of these, 29 are outliers on the BPT diagram \citep{Baldwin_1981} and can be classified via their location on the diagram. Another interesting type of galaxies in this group is the post-starburst (E+A) galaxy. The spectra of these 8 galaxies show H$\delta$ absorption tracing recent star formation, with sometimes still showing on-going star formation via the strong H$\alpha$ emission. At last, we have 4 galaxies showing either uncommon or unknown emission lines in their spectra. \\

\begin{figure}[ht!]
    \centering
    \includegraphics[width=0.55\linewidth]{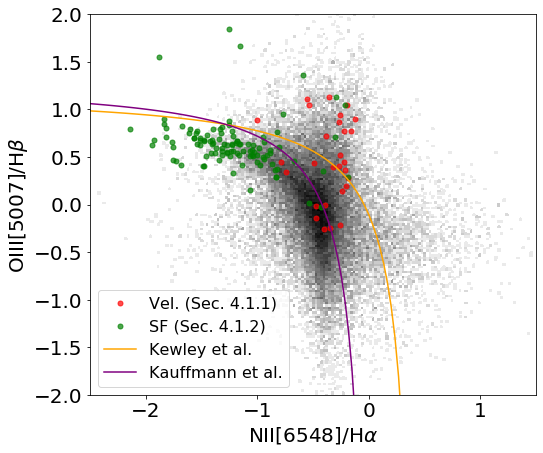}
    \caption{BPT diagram of the ratio NII[6548]/H$\alpha$ versus OIII/H$\beta$. Both the axes are in log space and the background is composed of the distribution of these ratios for all GAMA spectra. Two theoretical lines (\citealp{Kewley_2001} and \citealp{Kauffmann_2003}) probe the division between the star formation region and active galactic nuclei. The colored points are spectra from the groups described in Section \ref{sec:unusualvel} (red) and Section \ref{sec:emissoinlines} (green).}
    \label{fig:bpt}
\end{figure}

Many spectra show signs of star formation, as seen by the dominating strong Balmer emission lines present in the spectra. Additional to the Balmer lines, there are often many other emission lines. This often leads to looking like the continuum is a flat line, as can be seen in spectrum \texttt{G02\_Y5\_114\_291} in Figure \ref{fig:lines}. The heights and ratios of all these emission lines can be used to study the processes in the galaxies and trace the star formation phase or presence of active galaxies. We compute the ratios of a few emission lines and plot them on the BPT diagram. The BPT diagram is a popular method to classify emission line spectra as they can classify spectra very good based on simple observations. We use the provided emission line fluxes from the GAMA survey where available and compute the line emission ratios of NII/H$\alpha$ against OIII/H$\beta$. In Figure \ref{fig:bpt} we show the line emission ratios of all weird objects that belong to our kinematics and emission line groups. In the background, the distribution of the emission lines ratios of all GAMA spectra is shown as a comparison. For some objects, we have no emission line information as they fall outside of the observed wavelength range due to high redshift. 

\vfill

\clearpage

Many of the high emission line spectra, shown as green in Figure \ref{fig:bpt}, sit along the star formation track. The star formation track and AGN area in the BPT plot are divided by the purple \citep{Kauffmann_2003} and orange \citep{Kewley_2001} lines originating from earlier studies on galaxies. As the points in the BPT diagram indicate the classes we assigned to them, we can see that some of them are wrong as a few green points lie within the AGN region. This shows that some of the spectra we initially grouped by eye on the strong emission lines also have active processes in the galaxy. In total, there are 29 spectra that are on the extreme ends on in the BPT diagram and can be considered weird as they have very extreme line ratios. Only 10 of the spectra with strong emission lines are classified as either an emission-line Galaxy or HII Galaxy in the literature. \\

A special type of spectra not captured by the BPT diagram, but one with interesting emission line ratios, originates from port-starburst galaxies. These spectra have a strong H$\delta$ absorption line, which is an indication of an A-star dominated galaxy. These galaxies are often called E+A galaxies, as they originally thought that this type is only composed of Elliptical galaxies with A-type stars. Models suggest that this absorption is only possible for galaxies with a recent burst of star formation and now goes trough either none, or passive evolution of star formation \citep{Goto_2003}. We can differentiate two sub-types by looking at the presence of H$\alpha$ and Oxygen emission lines tracing ongoing star formation. We find 8 spectra with strong H$\delta$ absorption tracing recent a recent burst of star formation. Of these, 3 spectra show ongoing star formation, as emission lines are present. An example of an E+A galaxy, spectrum \texttt{G15\_Y4\_210\_008}, is shown in Figure \ref{fig:lines}.\\

Additionally to spectra with high emission lines, we also find 17 spectra with either weird line ratios or additional unknown lines. Of these 17 spectra, we have 10 spectra that lack any other emission line other than the H$\alpha$-NII structure. We also find 3 spectra that have complex NII emission structure dominating over H$\alpha$. Unfortunately, these spectra do not show up on the BPT diagram in Figure \ref{fig:bpt} as there is no flux information of these spectra due to the noisy additional structure. The last 4 spectra include uncommon or weird single emission lines and we look at these individually. \\

Spectrum \texttt{G09\_Y1\_CN2\_290} has numerous Iron lines along its broad emission lines. The Iron lines are present at 5303\angstrom, 5722\angstrom, 6088\angstrom, and 6376\angstrom\ as can be seen in Figure \ref{fig:lines}. These \textit{coronal lines} originate from gas exposed to X-ray radiation and is investigated in \citet{Wang_2012}. The Iron lines fade out over a time scale of years suggesting a transient nature. This is again a nice example of an observation of an event on a short time scale. \\

\begin{figure}[ht!]
    \centering
    \includegraphics[width=\linewidth]{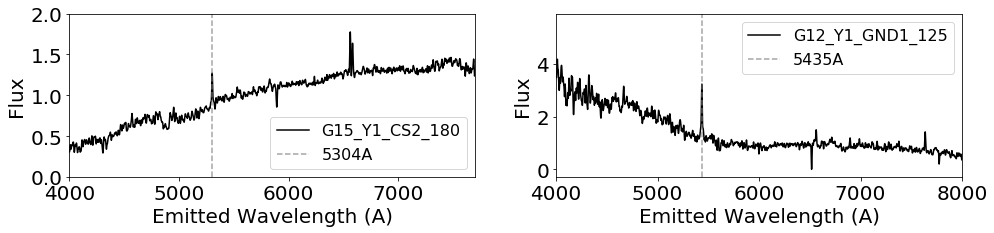}
    \caption{Two spectra with unusual single emission lines.}
    \label{fig:ironlines}
\end{figure}

In both the spectra \texttt{G15\_Y1\_CS2\_180} and \texttt{G12\_Y1\_GND1\_125} we observe a single unusual strong emission line which aligns with an Iron line. The spectra are shown in Figure \ref{fig:ironlines}, where the single emission lines sit at 5304\angstrom\ and 5435\angstrom\, tracing FeXIV and FeI respectively. In contrary to spectrum \texttt{G09\_Y1\_CN2\_290}, there is only a single Iron line suggesting either a different origin of the emission or absorption of the other emission lines in the line of sight. Also, the FeI line is very easy to ionize and will only be visible in very cool regions, making it a very weird observation. 

\clearpage

\vfill
The last spectrum with an uncommon line is \texttt{G12\_Y1\_HND1\_169}, which shows strong NI[5200] emission along the usual emission lines. This line is unusual, as atomic Nitrogen is easy to ionize and therefore not often observed. It can only exist in cool regions inside the galaxy shielded from ionizing sources. None of the spectra with unusual extra lines have been mentioned or investigated in the literature. As we primarily search for the outliers, we will not elaborate on the sources of these unusual lines.

\vfill

\subsubsection{Blends}
\begin{figure}[ht!]
    \centering
    \includegraphics[width=\linewidth]{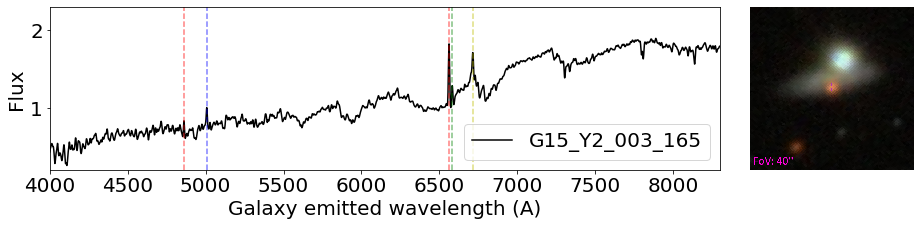}
    \caption{Example of an M-dwarf star in front of a galaxy. Note that we can see the characteristic star spectrum including emission lines of a galaxy. Balmer emission is indicated by red, OIII by blue, NII by green, and SII by yellow lines.}
    \label{fig:blend}
\end{figure}

We find 8 spectra that show features of two different objects. Almost all of the blends are composed of a foreground star in front of a galaxy. There are 6 spectra with an M-dwarf and 1 with an A-type star in front of a galaxy. The spectrum of an M-dwarf can easily be observed via its characteristic features. In Figure \ref{fig:blend} we show the spectrum \texttt{G15\_Y2\_003\_165} redshifted based on the emission lines of the background galaxy. In spectrum \texttt{G09\_Y5\_018\_163} (not shown here) an A-type star is found in front of a Galaxy which shows very blue emission, stellar absorption lines, and galactic emission lines. Visual inspection at the coordinates of the blended spectra give nice pictures of the stars in front of the galaxies. For the spectrum in Figure \ref{fig:blend}, the target object was actually the fore-ground star itself, as can be seen by the central marker in the image. At last, we also find 1 blended object in which a small galaxy is in front of a bigger galaxy. While the foreground galaxy is targeted, it is unknown if the emission lines originate from the foreground or background galaxy as the galaxies are almost at the same distance. None of the blended objects are found in the literature. 

\vfill

\clearpage

\begin{figure}[ht!]
    \centering
    \includegraphics[width=\linewidth]{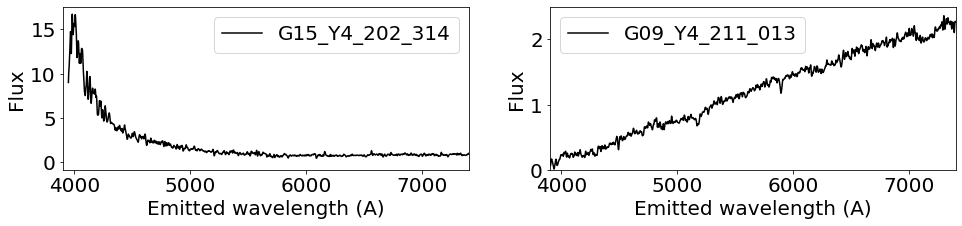}
    \caption{Two examples of galaxies we classify as extremely blue or diagonal red}
    \label{fig:continuum}
\end{figure}

\subsubsection{Continuum}
The last group of weird spectra is composed of (weird) features on a continuum level. This group mainly consists of extremely blue and diagonal red galaxies, as well as spectra with weird continuum shapes. In total, we have 33 spectra that probably have a high outliers score due to their continuum. Of these, 21 spectra have very high flux values in the blue end and 7 spectra are diagonal lines with high flux values at the red end of the spectra, as can be seen in Figure \ref{fig:continuum}. These two types have high outlier scores as the URF algorithm showed clustering based on a continuum level. This could also be seen in the visualizations of the sorted spectra on their weirdness scores in Figure \ref{fig:sortedall}. Along with these extreme galaxies, we also have 3 galaxies with high continuum flux at both the blue and red end making a sort of \textit{valley} shape. In contrary to this, we also have a \textit{hill} shaped spectrum with no emission lines present. \\

\begin{figure}[ht!]
    \centering
    \includegraphics[width=\linewidth]{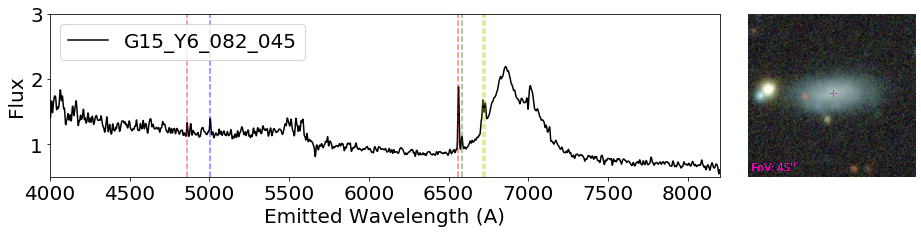}
    \caption{Spectrum with a seemingly weird velocity structure. We also show an SDSS DR9 picture of the galaxy, showing no extreme outflows.}
    \label{fig:045object}
\end{figure}

Most of these spectra are unwanted as outliers, as they are solely outliers due to their continuum shape without interesting features. However, we did find an object with a very weird structure which is shown in Figure \ref{fig:045object}. Spectrum \texttt{G15\_Y6\_082\_045} has a very big unexplained feature around 7000\angstrom. This spectrum had the highest outlier score in all of our runs and absorbed much of our time thinking about what this \textit{unknown unknown} could be. Sadly, this weird feature does not trace any physical process going on in the galaxy, as we eventually found three more spectra that show a very similar structure in the same position in the observed frame. Looking at those spectra combined, we also notice that this feature goes along with a bad splice. We classified the other three spectra as bad spectra, but keep \texttt{G15\_Y6\_082\_045} as an outlier as it could have been a very interesting find if it was real.

\vfill

\clearpage

\begin{figure}[ht!]
    \centering
    \includegraphics[width=0.6\linewidth]{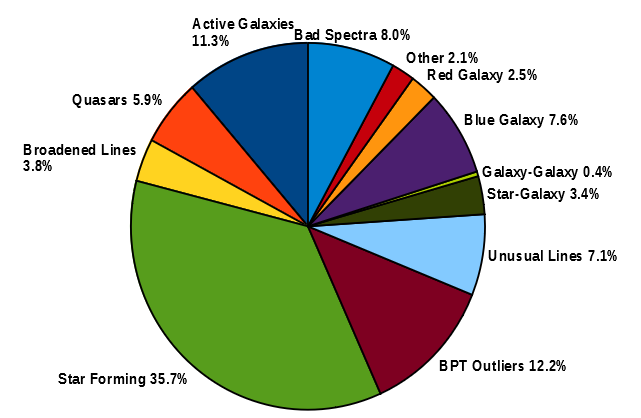}
    \caption{Overview of the types of outliers.}
    \label{fig:piechart}
\end{figure}

\vspace{0.5cm}

\subsubsection{Summary of Findings}
In our search of the weirdest galaxies in GAMA, we found many different types of galaxies. The overall findings are summarized in Figure \ref{fig:piechart}. Along with the real outliers, we also find 19 spectra that contain weird reduction errors or artifacts from reduced skylines causing a high outlier score. These spectra are different than the systematic bad spectra we flagged earlier and remain in the data set. Overall, the URF algorithm could find a lot of  outliers as shown in the previous sections. However, as can be seen in the pie chart and in the visualizations of the algorithm runs (see Fig. \ref{fig:sortedall}), we also find a few spectra that contain no interesting features. These spectra are clustered on their continuum as many spectra with either high blue flux values or high red flux values are found together. As this behavior was not inspected or mentioned in \citet{Baron_2016}, we do not know if this is a problem or feature of the algorithm itself or due to our implementation.

\vspace{0.5cm}

\subsection{Variational Autoencoder}\label{sec:vaeoutliers}
We also investigate the weirdest spectra found by the variational autoencoder. We will not investigate the top outliers as in-depth as done with the URF, but rather look at the performance of the variational autoencoder and the different types of outliers it can find. Therefor, we will not provide the tables of the weirdest spectra as done with the URF method, but refer the reader to the outlier scores provided on the Github page. We computed the outlier scores with the variational autoencoder using two different methods: via the reconstruction error, and by applying LocalOutlierFactor (LOF) on the latent space variables. Using the reconstruction based outlier scores we find 14 spectra within the weirdest 100 objects that were already found with the URF algorithm and were inspected in Section \ref{sec:urff}. For the LOF based scores, this is 15 spectra. Looking at the weirdest spectra from both the scores of the variational autoencoder, and the inspected spectra from the URF, we find 5 spectra that are outliers in all of the methods used in this research. Of these spectra, 2 have extremely high emission lines, and 3 show active processes.

\clearpage

\subsubsection{Reconstruction Error}
The most common way to use the variational autoencoder for outlier detection is by looking at the reconstruction error of the individual objects. Spectra containing either uncommon or weird features will not be reconstructed properly and its reconstruction error can be used as an outlier score. In Section \ref{sec:recon} we showed the reconstruction capability of the variational autoencoder on two types of spectra: blended spectra with a foreground star and galaxy, and spectra of different types of active processes in the galaxy. These examples showed that spectra with complex structure or emission at uncommon wavelengths will have a high reconstruction error, resulting in a high outlier score. We expect to find many spectra with these types of broad features, as those generate high errors. We also expect to find spectra with many strong or uncommon emission lines, as these can give a high outlier score via the squared relationship. \\

\begin{figure}[ht!]
    \centering
    \includegraphics[width=\linewidth]{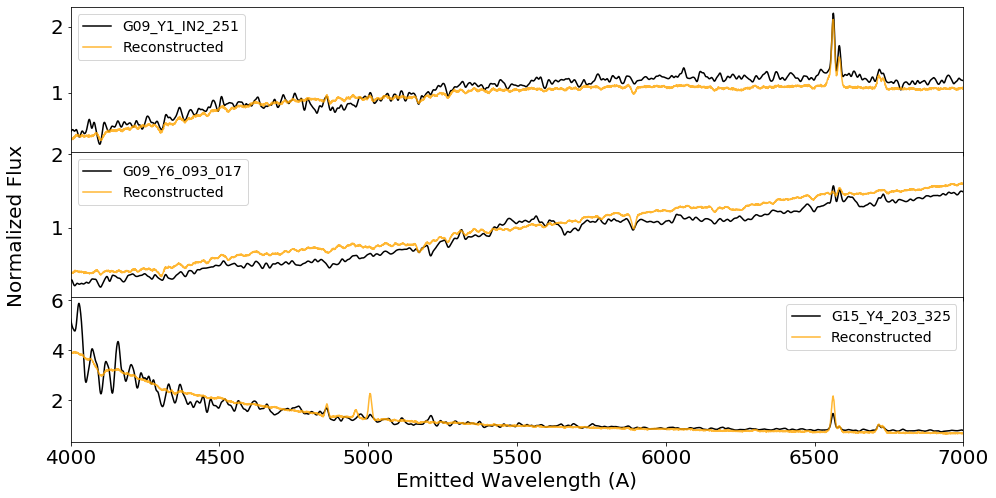}
    \caption{Examples of spectra with high reconstruction error primarily based on their continuum shapes. From top to bottom: Spectrum with a higher continuum at high wavelengths as reconstructed, a very nice edge-on spiral with more complex continuum as reconstructed, and noisy fluctuations at low wavelengths contributing significantly to the reconstruction score.}
    \label{fig:reconexamples}
\end{figure}

We inspect the 100 weirdest spectra with the highest reconstruction errors to look at the global performance of this outlier detection method. As we can inspect the reconstructed output of the variational autoencoder, this method makes it very easy to investigate why the input spectra are considered outliers. Of the outlying spectra, 64 spectra have a high outlier score due to either a weird continuum shape, or weird features originating from reduction errors. We show a few selected spectra in Figure \ref{fig:reconexamples} including their reconstructed version from the variational autoencoder. A (slightly) weird continuum shape can result in a high outlier score as almost all data points have a high reconstruction error. Among the reduction errors, we find many single high emission lines originating from inaccurate sky reduction or possible cosmic ray events. Also, the weird spectrum \texttt{G15\_Y6\_082\_045} (see Fig. \ref{fig:045object}) is again found as an outlier. Note that, while we also found many outliers based on their continuum with the URF method, the continua we find with the variational autoencoder always have very unusual shapes. The outlying continua found by the URF were mostly based on the clustering of very blue or red galaxies. 

\clearpage

\begin{figure}[ht!]
    \centering
    \includegraphics[width=\linewidth]{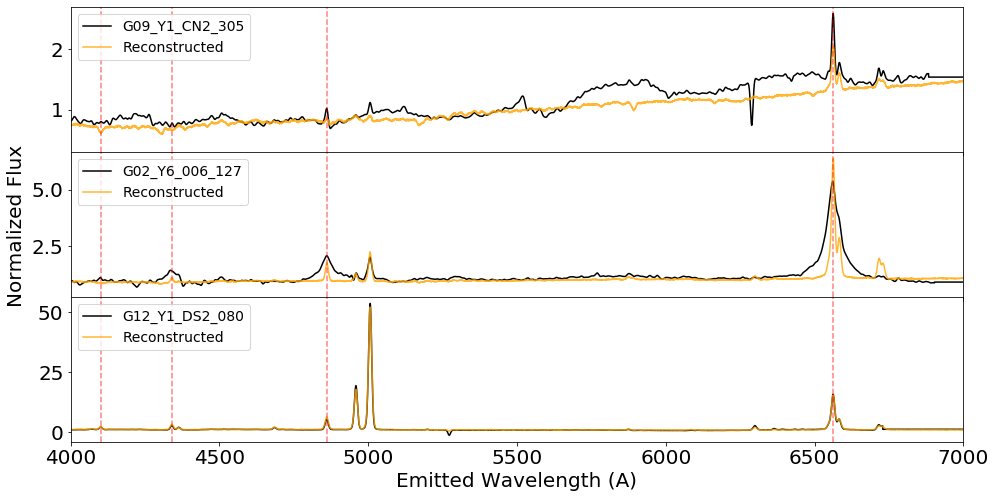}
    \caption{Selection of outliers found via the reconstruction error of the variational autoencoder, not found by the URF method. From top to bottom: blended spectrum containing an M-dwarf, AGN, and strong emission lines. Balmer line locations are indicated with the red line.}
    \label{fig:reconexamples2}
\end{figure}

The other 36 outliers, that contain no major reduction errors or unusual continuum shapes, are very similar to the types of spectra found by the URF. A selection of these outliers is shown in Figure \ref{fig:reconexamples2}. As expected, we find many spectra with high or broad emission lines, including spectra with complex structures at the emission lines. We also find two E+A galaxies, of which only one was found by the URF. At last, we find 8 blended objects consisting of an M-dwarf in front of a galaxy. \\

Overall, using the reconstruction errors as the outlier scores gives many weird spectra have unusual continuum shapes. This is expected, as for a weird continuum shape all of the data points contribute to the reconstruction error. We also find many interesting outliers, similar to the types of outliers found by the URF method. Using the reconstruction errors of the variational autoencoder as an outlier score can give good results, given that all the spectra with reduction errors are removed and weird continuum shapes are considered interesting.

\subsubsection{Latent Space Clustering}
The second method of outlier detection using the variational autoencoder is looking at the clustering of the latent space variables. This method completely omits the inference model of the variational autoencoder, using only the encoded latent variables to find outliers. Outlying spectra have either uncommon or weird features that are not recognized by the encoder, and might be encoded to a weird latent space vector. If we have found the outlying latent space vectors, then we have to use the inference model to decode the spectrum to try to understand why it is an outlier. So, while not using the inference model in the search of outliers, we still need it to investigate the outcome.

\clearpage

\begin{figure}[ht!]
    \centering
    \includegraphics[width=\linewidth]{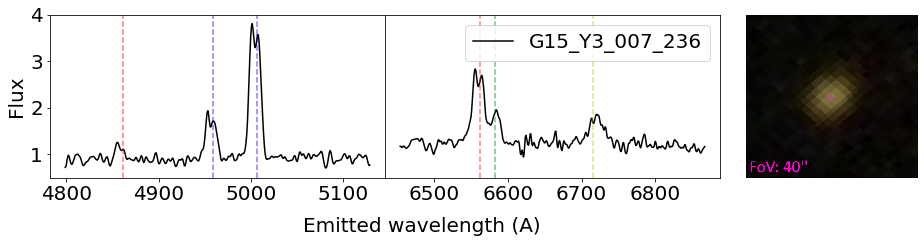}
    \caption{Spectrum showing double peaked emission lines. An extra blue-shifted component can be seen at the Oxygen emission lines (indicated by blue).}
    \label{fig:doubleline}
\end{figure}

We inspect the 100 weirdest latent space vectors by reconstructing them to spectra using the decoder. In comparison to the reconstruction errors as outlier scores, we quickly see that there is a lower amount of outliers based on their continuum or containing reduction errors. A total of 42 spectra are outliers due to this, while for 58 spectra we see other interesting features. The major difference between the reconstruction based and the latent based outlier scores is that we find more outliers based on their emission line strength and ratios. For example, there is not a single spectrum containing a blend of two objects, while we do find many spectra with weird emission line ratios. A few spectra show very strong OII emission relative to the other emission lines or strong NII emission in comparison to H$\alpha$. On top of this, we find a type of spectrum that we have not found before with any other method. As shown in Figure \ref{fig:doubleline}, spectrum \texttt{G15\_Y3\_007\_236} shows double peaked emission lines. The oxygen lines show an extra blue-shifted component, while the H$\alpha$ structure and SII emission lines are more complex than expected. We note, again, that the outlier scores of all spectra can be found on the Github page.\\

Comparing the outlier types found via the reconstruction error and the latent space, we note one fundamental difference. The reconstruction based method finds more outliers with an unusual continuum, as can be seen by the number of blends and weird continuum shapes that were found. This follows naturally from how the reconstruction error is defined, as all the data points contribute to the reconstruction error. Also, it might be that the encoder can not recognize the unusual continuum and maps it to random latent space variables, resulting in fewer spectra with an unusual continuum using the latent space approach. Depending on the types of outliers the user wants to find, both methods can be used to find interesting outliers.

\clearpage
\section{Visualization}\label{sec:tsne}
Data visualization is very important in either understanding or communicating complex data. Especially in big data projects with many objects, visualizing important features of all the data at once can be very complicated. The outlier detection algorithms we have used in this research generate high dimensional data that is not easily visualized. For example, the Unsupervised Random Forest algorithm produces a complex high dimensional distance matrix with the pair-wise distance between all objects. In \cite{Reis_2018a}, they produced a similar distance matrix of APOGEE spectra and visualized it using t-distributed Stochastic Neighbor Embedding (t-SNE, \citealp{Maaten_2008}). t-SNE reduces high dimensional data into a 2D map while preserving the distances and similarities between the objects. With the maps, they showed that the distance matrix contains information of different physical properties of the objects. \\

We visualize the high dimensional data from the URF and the variational autoencoder on a 2D map using t-SNE and inspect the clustering of similar objects. The distance matrix contains a lot of information about the objects and we expect to find many relations in the maps. For the variational autoencoder, the spectra are already mapped to the 6-dimensional latent space. These can also be mapped and visualized on a 2D map to inspect the clustering of similar spectra. \\

t-SNE is a non-linear dimensionality reduction algorithm capable of reducing high dimensional data into two or three dimensions for visualization. We will reduce our data to two dimensions and visualize the results on a map. The parameter space of the map itself does not trace any physical properties, but similar objects are found in clusters on the map. t-SNE tries to preserve the nearest-neighbors of each point, while forcing them to a lower dimension. Object far away from each other are less similar, but there is no relationship between the absolute distance between the points and their dissimilarity. The first step of the t-SNE algorithm is constructing a pair-wise probability distribution indicating the similarity between objects. The second step is to define a similar distribution of points in a two or three-dimensional map and minimize the KL divergence between the distributions. The KL divergence was mentioned and explained in Section \ref{sec:neuralnetwork}.\\

We apply t-SNE on our data using the Python implementation in the scikit-learn\footnote{\url{https://scikit-learn.org/stable/modules/generated/sklearn.manifold.TSNE.html}} package. Two important hyperparameters are the \textit{perplexity} and the \textit{learning rate}. The perplexity is related to the amount of nearest-neighbors used as similar objects. This controls the size of clusters on the final mapped representation of the input. A higher perplexity results in more global clustering, while low perplexity shows clusters on smaller scales. This parameter is very sensitive and has a big influence on the results. The learning rate is, as in any other machine learning application, important for the convergence of the algorithm. If the learning rate is set to high, then the final map will look like a ball in which all points have the same distance from each other. If the learning rate is set to low most of the points are found in dense clusters with a few outliers. In our application, we tried different values for these parameters and we will show the best visualization of our data. The best visualization is a t-SNE map that shows clustering of similar objects on both a global, and a local scale. We try to not have any dense regions due to overfitting, or have any significant sparse region without data points. For the distance matrix, we get a map that fits these requirements using a perplexity of 200 and a learning rate value of 5000. 

\clearpage

\begin{figure}[ht!]
    \centering
    \includegraphics[width=\linewidth]{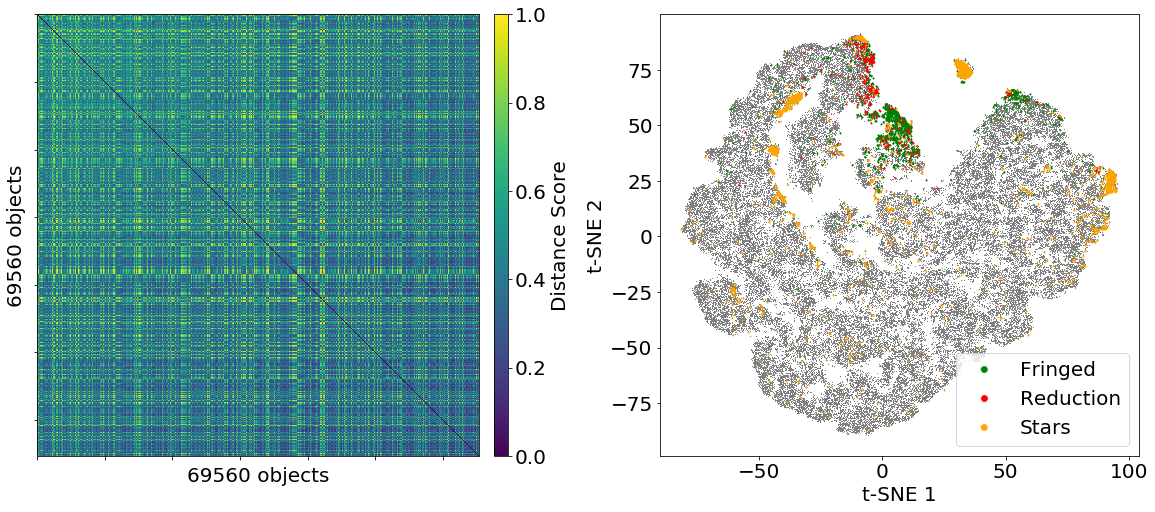}
    \caption{\textit{Left:} distance matrix from all objects in \texttt{SNR5} subset (including bad spectra and stars). The objects are sorted on signal to noise ratio, so we can even see the bias of the URF algorithm as higher distance scores are shown in the top left corner. \textit{Right:} t-SNE representation of the distance matrix including flagged objects. Note the grouping of similar objects.}
    \label{fig:distancetsne}
\end{figure}

\subsection{Distance Matrix}
The URF outlier detection method produces a distance matrix with the pair-wise distance between each object. We used this distance matrix to compute the distance (weirdness) score of each individual object and inspect the weirdest objects. We will reduce the distance matrix of the \texttt{SNR5} subset to a 2D map and inspect the clustering of similar objects. The \texttt{SNR5} subset is composed of good signal to noise ratio spectra, and its big size makes it a nice data set for the visualization.\\

The t-SNE map of the distance matrix is shown in Figure \ref{fig:distancetsne}, which represents the full \texttt{SNR5} subset including bad spectra and stars. As stated earlier in Section \ref{sec:urf}, the URF algorithm assigns higher distance scores to lower signal to noise ratio galaxies. This can be seen in the distance matrix where higher distance scores are assigned to the lower SNR galaxies in the top-left corner. In the t-SNE map in Figure \ref{fig:distancetsne} three groups of spectra are marked. These are the fringed spectra, spectra with reduction errors, and the stars in the GAMA survey. We included these spectra to show that the URF algorithm is capable of grouping similar types of galaxies based on their common features.\\

\begin{figure}[ht!]
    \centering
    \includegraphics[width=\linewidth]{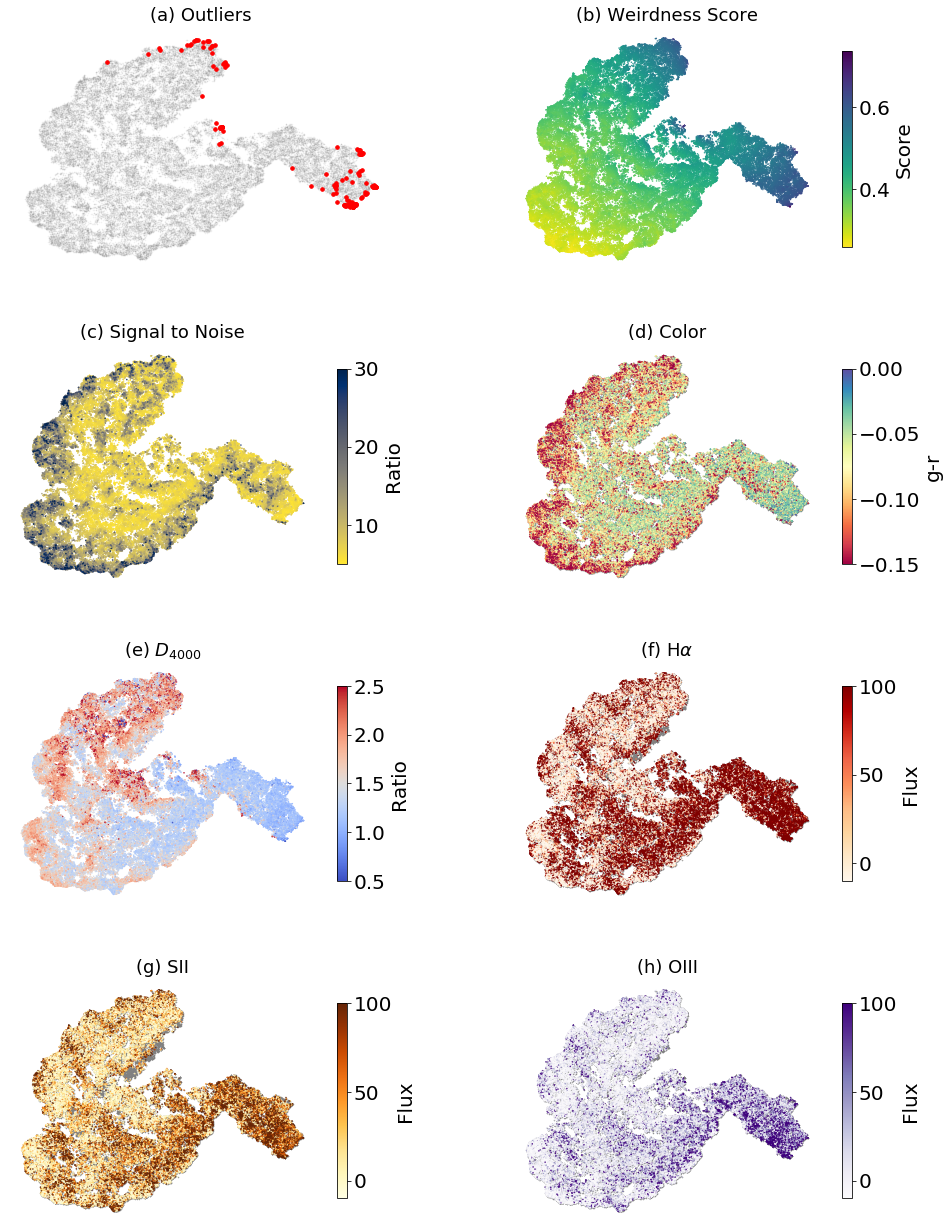}
    \caption{Mapped distribution of the distance matrix via t-SNE. The objects are mapped onto a two-dimensional plane in which similar objects are clustered locally. Different data parameters or galaxy feature data is shown as color maps to indicate the amount of information the distance matrix contains.}
    \label{fig:distancetsne2}
\end{figure}

For the best results, we flagged the bad spectra and stars during this research to ensure that they do not interfere with learning the features of good spectra. This resulted in the removal of 5667 spectra from the \texttt{SNR5} subset, giving a slightly smaller distance matrix after a new run with the URF algorithm. We use t-SNE on the new distance matrix to compute the map of only good spectra. As the dimension of the distance matrix has been lowered, the new t-SNE map will look different as the map shown in Figure \ref{fig:distancetsne}. The new map is shown multiple times in Figure \ref{fig:distancetsne2} with features of the objects to inspect the clustering in the map and the correlations between different parameters and the t-SNE space.

\clearpage

\clearpage

In Figure \ref{fig:distancetsne2}, we show the t-SNE map of the \texttt{SNR5} subset with the following parameters as color map, from top-left to bottom-right: the 250 outliers, the weirdness score, the signal to noise ratio of spectra, the color of objects, D4000 break index, H$\alpha$ line emission, SII line emission, and OIII line emission. While the signal to noise ratio is not a physical parameter of the objects, it is included to show the correlation between the weirdness score and its quality. The color maps are chosen as they represent the characteristics and features of the spectra. As the distance matrix is reduced from a high dimension to only two dimensions, much hidden information is lost. Still, we find parameters that show structure on the map and we comment on each of them and explain the findings.\\

The weirdness score overview in map (b) shows three regions in which the scores are the highest. These regions can also be traced back by the highlighted outliers in map (a), in which the marked points are the 250 outliers that were inspected in Section \ref{sec:outlier}. The outliers are found in different places in the t-SNE map with big distances in-between them, suggesting that they are composed of different types of outliers. This corresponds to the different groups of outliers we found and similar outliers in these groups are found in small clusters on the t-SNE map. We also included a color map of the signal to noise ratio of each object in map (c). The higher signal to noise spectra are found at the outskirts of the map. When looking at map (b) and (c), we can see the correlation between the low signal to noise spectra and high outliers scores. In map (d) the color of the objects are shown using the SDSS \textit{g} and \textit{r}-band photometry. Galaxies with similar color are also grouped along the edge of the t-SNE map. There is also a relation between the high signal to noise ratio and the color. Red galaxies tend to have an overall higher signal to noise ratio, as the signal is better at higher wavelengths relative to lower wavelengths.\\

At last, we have four maps based on the features in the spectra. These maps trace the physical properties of the spectra and can be used to trace the information contained in the distance matrix. The $D_{4000}$ break index is used to determine the star formation characteristics of galaxies. It is defined as the flux at the wavelengths 3750\angstrom-3950\angstrom\ divided by the flux at wavelengths 4050\angstrom-4250\angstrom\ \citep{Bruzual_1983}. These regions trace many lines that correspond to either young bright stars or an older population of metal-rich stars. An interesting observation is that while there is a clear distinction between the two types of galaxies, for both types we have a region with high weirdness scores and objects marked as outliers as seen in (a). The following three maps (f), (g), and (h) show the flux values of the emission lines H$\alpha$, SII, and OIII respectively. These lines are present during star formation and a young star population. Note that for the H$\alpha$ line and OIII line, the regions of the highest intensities do not always correspond to each other. They do overlap in the rightmost part of the map indicating the star formation region.\\

The clustering of the groups of outliers indicates that there are more similar objects nearby. Inspecting the nearby points indeed gives us more similar objects. The region with star-forming galaxies is pretty global and easily spotted by the flux maps. More star-forming galaxies are easily found by inspecting points close to the marked outliers. The BPT outliers are clustered at the tip of the star formation region. The quasars and active galaxies can not be found on a global scale and reside in a small group. More interesting spectra containing very broad emission lines or extra structure could be found by inspecting the points close to our outliers with unusual velocity structure. Overall, the t-SNE map of the distance matrix is very useful in finding more interesting spectra, while also looking very pretty.

\clearpage

\subsection{Latent Variables}
\begin{figure}[ht!]
    \centering
    \includegraphics[width=\linewidth]{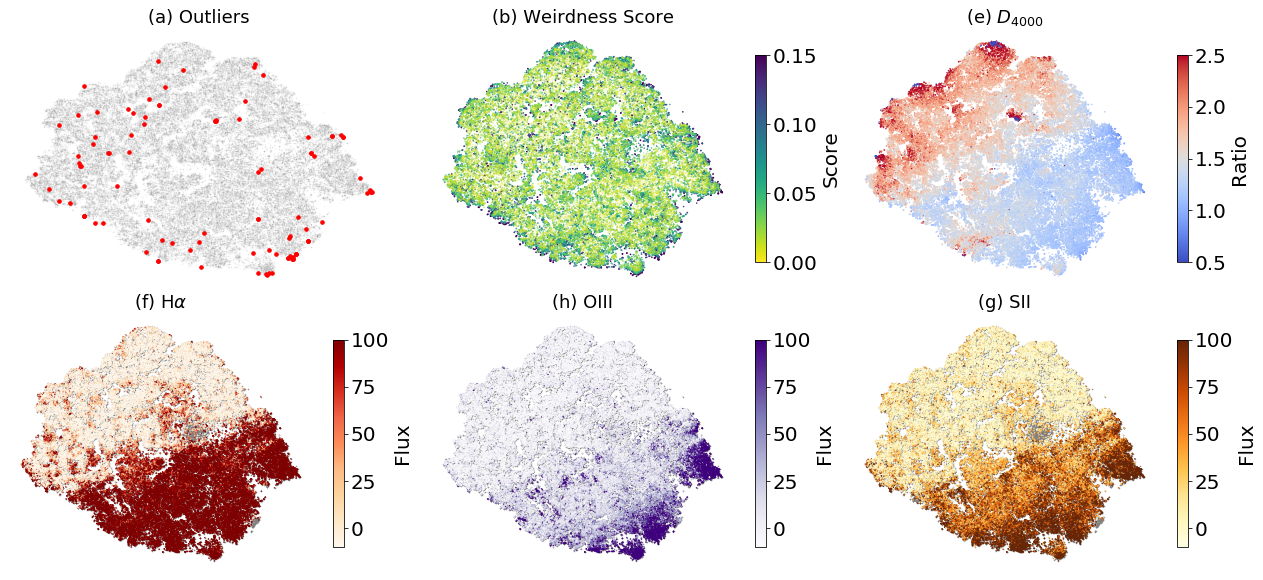}
    \caption{Mapped distribution of the latent vectors via t-SNE. Different data parameters or galaxy feature data is used as color maps. We do not show (c) and (d), as the objects are the same as already shown in Figure \ref{fig:distancetsne2}.}
    \label{fig:tsne3}
\end{figure}

We also want to test the dimensionality reduction of the 6 latent variables representing the encoded spectra. In \citet{Portillo_2020}, the latent variables were plotted using a corner plot, showing the pair-wise relationships between them. Instead, we will use the latent space vectors in a similar way as the distance matrix of the URF and try to make a nice t-SNE map. As we could find interesting outliers with the latent space vectors, we expect to find a good map in which we can find more interesting outliers. We apply t-SNE on the encoded spectra and used different learning rates and perplexity values to find the best map. The best t-SNE map we found is shown in Figure \ref{fig:tsne3}, using a perplexity of 200 and learning rate value of 1000.\\

Overall, we see the same types of relations as shown for the distance matrix in Figure \ref{fig:distancetsne2}. The outliers, as shown in (a), seem to be at random places, but are mostly found at the edges of the whole map or at the edges of dense regions in the map. This can also be seen by the LOF outlier scores of the spectra in (b). The flux values of the emission lines H$\alpha$, OIII and SII are shown in (f), (h) and (g) respectively. The objects with high flux values for these emission lines are mostly found in the bottom-right corner of the map, tracing the star formation region of the t-SNE map. \\

Many similar objects can be found in the star-forming region of the distance map, which was also possible with the map in Figure \ref{fig:distancetsne2}. Unfortunately, we could not find any more double peaked emission lines in the neighborhood of the found spectrum. The maps show a very nice global distribution and clusters of the different parameters, but the 6 latent parameters contain less information about the spectra than the distance matrix, as the encoder and decoder also contain much information about relations in the data. This gives that there is almost no clustering of similar objects on a small scale. The best t-SNE map is made with the very detailed distance matrix, but Figure \ref{fig:tsne3} shows that the latent variables also nicely trace features and cluster similar objects on a larger scale. 

\clearpage
\section{Discussion}\label{sec:discussion}
In this section, we discuss the decisions made during the project and interpret our results. We comment on the acquisition of the GAMA data and the pre-processing steps done on the spectra. We discuss our application of the URF outlier detection method and compare the results with \citet{Baron_2016}. Also, the choices made on defining our outlier groups are discussed, along with comparing the different outliers found between the two detection methods. At last, we comment on the t-SNE maps and discuss the application of all our methods for future projects.\\

While using a major part of the GAMA survey, we did not use all spectra of the survey. We limited ourselves to only use the data of the main observations originating from a single instrument, making up 78\% of the spectra. The decision was made for the following two reasons. At first, all spectra are taken with the same instrument and have the same resolution and overall data shapes. If data from multiple instruments were used, the outlier detection methods could be biased towards the data of a single instrument. Secondly, we only have to work with the instrumental errors of a single instrument, making masking of bad spectra easier.\\

The GAMA survey is composed of objects at a broad range of redshifts. To learn the important features, the spectra had to be converted to their rest-frame such that all emission lines are aligned. To ensure the same input space of each spectrum, the spectra were interpolated with the rest frame wavelengths between 3500\angstrom\ and 7500\angstrom. This ensured that the interesting emission lines are captured for all the objects at different redshifts and excluded regions without flux values. In an ideal world, we would use the raw flux values of the spectra for the outliers search, as they ensure the best relationship between what is observed and an assigned score. However, many spectra showed either bad regions or noisy parts that had to be accounted for. Therefore we masked bad regions and smoothed the spectra with a Gaussian kernel. The fluctuations in the continuum were removed successfully and the emission line shapes stayed intact, but the absolute flux values were slightly lowered. This had no impact on the results, as all fluxes were lowered with the same ratio. We think we made a good trade-off between the pre-processing of the data and keeping it as close to the observed values.\\

We implemented the full URF algorithm in Python using multiprocessing to boost its performance. A basic example of the algorithm was provided in \citet{Baron_2016}, in which they described simple steps to build the algorithm. While the URF outlier detection method was already proven to be working in their work, we also wanted to explore how the algorithm works. Therefore, we applied different tests by changing the input shape of the data, using different hyperparameters in the Random Forest, and inserting self-made spectra that look very weird. The tests showed that spectra with broad features are easily found as outliers, while unusual single lines do not necessarily show up as outliers. Overall, the tests gave good insight into the processing of the data and the application on the URF.\\

Initial runs on all our GAMA data only showed outliers due to very noisy fluctuations. We observed that the URF algorithm is biased towards low signal to noise ratio spectra, which was coincidentally confirmed during this project by \citet{Reis_2019}. To ensure that we find the best outliers, we used three different subsets based on the signal to noise ratio of the spectra. While the subset with the high signal to noise spectra showed the best results, we also had to use the lower quality spectra to ensure we find the weirdest spectra in the GAMA survey. As we ran the algorithm on different subsets based on a lower limit of their signal to noise ratio, we had to carefully define which spectra are considered \textit{weird}. For the three subsets \texttt{SNR10}, \texttt{SNR5} and \texttt{SNR2.5}, in which the number indicates the lower limit of the signal to noise ratio, we defined the weirdest spectra as the 100 highest scoring spectra of \texttt{SNR10}, the 100 highest scoring spectra of \texttt{SNR5} not found earlier, and the 50 highest scores spectra of \texttt{SNR2.5} not found earlier. While this seems arbitrary, this allowed us to learn the workings of the algorithm, as we first applied good quality data to test the algorithm. Additionally, we only used 50 spectra of the \texttt{SNR2.5} subset, as the results were not great due to bias towards lower signal to noise galaxies. We also noticed clustering on the continuum of spectra in the outlier scores as many similar blue or red spectra are grouped together. These also came up as the weirdest spectra, but were in itself not very interesting. This behavior was not reported in \citet{Baron_2016}, while \cite{Reis_2019} shows similar behavior in the score distribution. Overall, interesting weird spectra still dominated the results, and we could still inspect spectra of many different types of objects.\\

The inspected outliers are grouped and categorized by their most obvious features. Note that, while we inspected the weirdest spectra thoroughly, we can not guarantee that we always noticed the exact reason why the URF algorithm assigned a high outlier score. As many features are obvious to spot by eye, it is entirely possible that the URF found a weird combination of flux values that determined its high score. We are however very confident that we grouped and classified the 250 weirdest spectra correctly. We provide a full list of the weirdest spectra on the Github page for the reader to explore themselves. We found many different types of spectra among the outliers, such as: additional complex structure at emission lines, line broadening due to activity, blends of different objects, and BPT outliers. While the tests showed that the algorithm does not assign high outlier scores to galaxies with unusual lines, we still find a few spectra with unusual Iron lines and NI[5200] emission. Many types of outliers we found are similar to the outliers found in \citet{Baron_2016}, but unfortunately there was no overlap of objects in our GAMA data and their SDSS data, so no direct comparisons could be made. As most weird spectra showed interesting features, we have great confidence in the application of the URF outlier detection method on spectroscopic data. One could argue that for all the weird spectra we found, simple models specific for each type of object could be used instead. While this is true for known objects, the outlier detection methods can find all types of objects, including unknowns. Unfortunately, we did not find any Nobel prize winning \textit{unknown unknown} in our research. \\

We also applied a more experimental outlier detection method on the galaxy spectra of the GAMA data using a reconstruction-based variational autoencoder to inspect more outliers and learn about the application of neural networks on this type of data. Using a reconstruction based method would make it easier for us to investigate why a spectrum is considered an outlier. We applied our self-made algorithm in a similar method as in \citet{Portillo_2020}, but with the primary goal of finding outliers. During training, we noticed that the \textit{MMD} loss term, which is supposed to help the network with learning, did not contribute enough. At first, we did not include a scaling term in the loss function that uses the batch size as a multiplier. However, when we did implement it correctly the results worsened significantly and were unusable. We could not find why this was happening and applied the InfoVAE as a normal variational autoencoder. We used the trained network to compute outlier scores for the spectra in the \texttt{SNR5} subset via two methods: the conventional reconstruction error, and via clustering of the latent space variables. We could find interesting outliers with both methods, but also found many spectra with a weird continuum shape or weird features due to reduction errors. We used a modified reconstruction metric as was found in the literature to normalize the scores between 0 and 1. To improve these results, the flux errors could be included in the reconstruction metric, or a completely different metric could be used. Due to time constraints, we have not tested a different metric. \\

We expected that the results would be very different between the URF and variational autoencoder techniques as they are completely different methods. However, we found that both methods find very similar types of outliers, and even had overlapping results. In terms of interesting outliers, the URF did perform better, while the variational autoencoder had some bias towards continuum based outliers. Still, the experimental reconstruction based method showed good results, and can possibly be improved with more training and research into the hyperparameter optimization of a bigger, more complex, network. \\

The visualization of the distance matrix with t-SNE showed good results, as the spectra in the different groups of outliers were all clustered on the map. Also, we could find more similar spectra by inspecting nearby points in the maps. While the t-SNE parameter space itself does not have any physical meaning, we could find many correlations with the features of the objects. Visualization of complex data is very handy and can be used to explore more data and find more interesting objects. Our results are similar to \citet{Reis_2018a}, in which they applied t-SNE on the distance matrix of stellar spectra. During our project, they published an interactive page \citep{reis_2019effectively} using a different dimensional reduction technique, where the user can inspect the parameter space of SDSS galaxy spectra. This interactive page can be used to generate similar maps as we made in Section \ref{sec:tsne}, and is a very nice application of data visualization. We also applied t-SNE to the latent space vectors of the encoded spectra from the variational autoencoder. The results were very similar, as the latent space variables could also trace the features in the spectra. The latent space variables do not make up all of the information of the spectra, as the encoder and decoder are also composed of many relations of the data. Therefore, the t-SNE maps using the distance matrix contain more information and have better local clustering of similar objects.\\

The outlier detection methods we used show great potential to cope with many spectroscopic observations of objects. These methods will be very useful for future surveys that generate thousands of spectra. Combining the outlier detection algorithms and good visualization tools helps enormously with either finding interesting objects or disregarding common "boring" objects. While we had a few issues with the URF with a bias towards low signal to noise ratio spectra and clustering on the continuum, the distance matrix still traced many physical parameters and gave interesting outliers. We also did a very informative experimental application of the variational autoencoder on spectral data to find outliers, and think that even with the brief overview of the results, it works very well.

\clearpage
\section{Summary and Conclusion}
We applied two different outlier detection methods on spectroscopic data from the GAMA survey to find weird galaxies in the Universe. The first method was a distance-based outlier detection technique based on an Unsupervised Random Forest. This method computes a pair-wise distance between all objects resulting in a high-dimensional distance matrix. We used the distance matrix to determine an outlier score for each individual object and inspected the 250 weirdest galaxies. We found many different types of interesting outliers, such as: quasi-stellar objects, active galaxies, blends of different objects, and BPT outliers. The method also assigned high outlier scores to objects clustered on their continuum and we observed a bias towards low signal to noise ratio spectra, which was also confirmed by others during the span of this project. Both these issues did not have any significant impact on the quality of the outliers. The second method is a reconstruction-based outlier detection technique using a variational autoencoder. The network was trained using the spectra and learns to reconstruct the important and common features. Spectra with weird or uncommon features will have a high reconstruction error and are considered outliers. We used the reconstruction errors, and the encoded representation of the spectra, to compute outlier scores and compared the outcome with the Unsupervised Random Forest. While the methods are inherently very different, they both found similar types of outliers and even had some overlap in the weirdest spectra.\\

We visualized the high dimensional output of the algorithms on a two-dimensional map using the dimensionality reduction technique t-SNE. This gave a great overview of all the information contained in the output and gave a nice insight into the relations found in the data. We could point out multiple correlations in the output using galaxy feature data. The maps could also be used to find even more interesting outliers by looking at points clustered on the map.\\

In this project, we successfully showed the application of two outlier detection techniques for spectroscopic data. These will be very useful for future surveys in which a lot of objects will be observed. Combining these methods with good visualization techniques gives a very robust way of finding many interesting objects. While we did not find the weirdest galaxy in the Universe in this project, we found many different interesting objects via a single method without the need for prior knowledge of any of them. Outlier detection is a very interesting application of unsupervised machine learning, and will be essential for finding the \textit{unknown unknowns} of the future.

\clearpage
\footnotesize
\bibliography{biblio.bib}
\bibliographystyle{mnras}
\normalsize

\clearpage
\section*{Acknowledgements}
I would like to thank my supervisors Teymoor Saifollahi (MSc) and prof. dr. Reynier Peletier for providing and helping me with this project. This master research project was a great way of combining the interesting field of Astronomy with the fast developing field of data science and machine learning. This project allowed me to expand my knowledge of the application of machine learning techniques on complex data and I am grateful that this is possible at the Kapteyn Institute. I want to thank my supervisors for their feedback, and especially thank Teymoor for our weekly (online) meetings where we shared and discussed our findings.\\

Also, I would like to thank prof. dr. Michael Biehl and dr. Kerstin Bunte for their knowledgeable input at the start of the project about data pre-processing and application of the machine learning techniques. At last, I also want to thank my girlfriend, parents, sister and friends for their \textit{gezelligheid}, discussions and support during the project. 
\clearpage
\section*{Appendix A: GAMA Data}\label{sec:dataacq}

SQL query for \url{http://www.gama-survey.org/dr3/query/}
\begin{verbatim}
SELECT SPECID, RA, DEC, Z, NQ, PROB, URL, URL_IMG, CATAID,
       GAMA_NAME, IC_FLAG, DIST
FROM   SpecAll
WHERE  SpecAll.NQ >= 3 
AND    SpecAll.SURVEY_CODE = 5 
AND    SpecAll.IS_SBEST = 1 
\end{verbatim}



\section*{Appendix B: Evidence Lower Bound in Practice}
We shortly relate the theory of the Evidence Lower Bound (ELBO) function contents to their application in the code of the variational autoencoder. The following functions and theory is shortly summarized out of \citet{Zhao_2017}, \citet{Ichinohe_2019} and \citet{Kingma_2019}.

\subsection*{Kullback-Leibler Divergence}
The \textit{Kullback-Leibler Divergence} (KL Divergence) assigns a distance between the approximate posterior and true posterior, and determines the distance between the ELBO and the marginal likelihood. As the prior is set to be the normal distribution $p(\bm{z}) = \mathcal{N}(0, I)$, the KL Divergence is:
\begin{equation}
    D_{KL} (q(\bm{z}|\bm{x}^i)||p(\bm{z})) = - \frac{1}{2}  \sum_{j=1}^J \Big(( 1 + \log\left[ (\sigma_j^i)^2  \right] - (\mu_j^i)^2 - (\sigma_j^i)^2  \Big)\ ,
\end{equation}
where $\mu_j^i$ and $\sigma_j^i$ are the \textit{j}th element of the mean $\bm{\mu}$ and variance $\bm{\sigma}$ calculated in their corresponding layer for each spectrum \textit{i} in the data set.
For reference, the variable \textit{J} denotes the amount of latent space variables and is set to $J = 6$ in our variational autoencoder.

\subsection*{Maximum-Mean Discrepancy}

The \textit{Maximum-Mean Discrepancy} (MMD) computes the distance between two probabilities by comparing their momentum. It is efficiently implemented using a kernel trick with \textit{k} any positive definite kernel. The MMD between the output $q(z_\phi)$ of the encoder and the prior $p(z_\theta)$ of the decoder is
\begin{equation}
    D_\text{MMD}(q(\bm{z}) || p(\bm{z})) = \mathbb{E}_{p(\bm{z}),p(\bm{z}')}[k(\bm{z},\bm{z}')] +
                                 \mathbb{E}_{q(\bm{z}),q(\bm{z}')}[k(\bm{z},\bm{z}')] - 
                                 2\mathbb{E}_{q(\bm{z}),p(\bm{z}')}[k(\bm{z},\bm{z}')]\ ,
\end{equation}
with $\mathbb{E}$ the expectation value of the output of kernel \textit{k}. We use a Gaussian kernel that computes the similarity of the two samples as in
\begin{equation}
    k(\bm{z}, \bm{z}') = \exp\left( -\frac{|\bm{z}-\bm{z}'|^2}{2\sigma^2} \right)
\end{equation}
The MMD loss is computed in our variational autoencoder by comparing the momentum of the sampled latent variable layer (z\_sampler) with the normal distribution $\mathcal{N}(0, I)$ of the prior.

\clearpage
\section*{Appendix C: Variational Autoencoder}

\begin{figure}[ht!]
    \centering
    \includegraphics[width=0.73\linewidth]{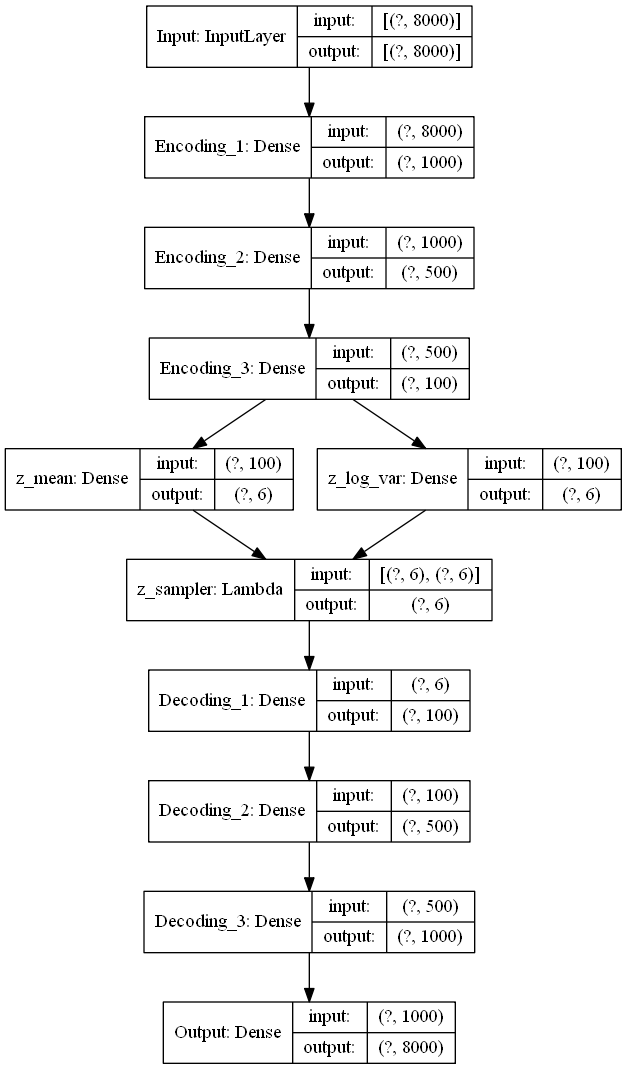}
    \caption{The layers and their sizes of our optimal variational autoencoder with input size 8000. For the network with input size of 1000, the network looks similar with the first hidden layer as the input layer. The question marks in the layer sizes indicate the batch size, which can be varied.}
    \label{fig:neuralnet}
\end{figure}
\vfill

\clearpage
\section*{Appendix D: Weirdest Galaxies in \st{the Universe} GAMA}
In the following tables we list the weirdest spectra we found with the URF algorithm. Additional information is shown that is relevant for each group and comments are provided that determined the placement of the spectra in their particular group. 

\tiny
\centering
\begin{tabularx}{\linewidth}{l|c|X}
\hline
SPECID & z & Comments \hspace{11cm} \\
\hline
G02\_Y5\_083\_265 & 0.185 & M dwarf in front of Galaxy               \\
G09\_Y1\_AX2\_002 & 0.006 & M dwarf in front of Galaxy               \\
G12\_Y1\_DS2\_320 & 0.179 & M dwarf in front of Galaxy               \\
G15\_Y1\_BS2\_010 & 0.357 & M dwarf in front of Galaxy               \\
G15\_Y2\_003\_165 & 0.049 & M dwarf in front of Galaxy               \\
G15\_Y2\_013\_324 & 0.082 & M dwarf in front of Galaxy               \\
G15\_Y3\_021\_053 & 0.235 & M dwarf in front of Galaxy               \\
G09\_Y5\_018\_163 & 0.040 & Bright star in front of Galaxy          \\
G09\_Y1\_HS1\_004 & 0.018 & Tiny galaxy in front of big one, H$\alpha$ source unknown\\
\hline
\caption{Blended objects}\label{tab:blends}
\end{tabularx}

\tiny
\centering
\begin{tabularx}{\linewidth}{l|c|X}
\hline
SPECID & z & Comments \hspace{11cm} \\
\hline
G09\_Y1\_FS1\_085 & 0.199 & Broadening of emission lines        \\
G09\_Y4\_212\_273 & 0.158 & Broadening of emission lines        \\
G09\_Y6\_063\_013 & 0.334 & Broadening of emission lines        \\
G12\_Y1\_DS2\_080 & 0.306 & Broadening of emission lines        \\
G12\_Y1\_IS1\_080 & 0.282 & Broadening of emission lines        \\
G15\_Y2\_007\_169 & 0.330 & Broadening of emission lines        \\
G12\_Y1\_BX1\_127 & 0.153 & Complex and high NII                \\
G02\_Y3\_015\_210 & 0.043 & H$\alpha$ and NII are complex lines \\
G15\_Y4\_206\_189 & 0.180 & H$\alpha$ and NII are complex lines \\
\hline
\caption{Minor broadening of, or minor additional structure in, emission lines}
\label{tab:broadening}
\end{tabularx}

\tiny
\centering
\begin{tabularx}{\linewidth}{l|c|X}
\hline
SPECID & z & Comments \hspace{11cm} \\
\hline
G02\_Y5\_080\_045 & 0.075 & Additional structure at Balmer lines     \\
G12\_Y1\_BN1\_366 & 0.243 & Additional structure at Balmer lines     \\
G15\_Y1\_BN2\_004 & 0.270 & Additional structure at Balmer lines     \\
G15\_Y6\_093\_351 & 0.317 & Additional structure at Balmer lines     \\
G09\_Y6\_090\_043 & 0.549 & Additional structure at H$\beta$ and broad MgII \\
G02\_Y3\_005\_005 & 0.330 & Broadened Balmer lines                   \\
G02\_Y3\_015\_281 & 0.325 & Broadened Balmer lines                   \\
G02\_Y5\_094\_233 & 0.157 & Broadened Balmer lines                   \\
G02\_Y5\_111\_257 & 0.331 & Broadened Balmer lines                   \\
G09\_Y1\_BS1\_152 & 0.176 & Broadened Balmer lines                   \\
G09\_Y4\_208\_024 & 0.295 & Broadened Balmer lines                   \\
G12\_Y6\_040\_320 & 0.106 & Broadened Balmer lines                   \\
G15\_Y1\_FN2\_029 & 0.331 & Broadened Balmer lines                   \\
G15\_Y1\_GN1\_046 & 0.144 & Broadened Balmer lines                   \\
G15\_Y4\_236\_287 & 0.323 & Broadened Balmer lines                   \\
G15\_Y5\_003\_263 & 0.300 & Broadened Balmer lines                   \\
G15\_Y5\_029\_255 & 0.279 & Broadened Balmer lines                   \\
G15\_Y5\_061\_182 & 0.220 & Broadened Balmer lines                   \\
G15\_Y6\_088\_158 & 0.140 & Broadened Balmer lines                   \\
G15\_Y6\_100\_268 & 0.235 & Broadened H$\alpha$                             \\
G09\_Y3\_015\_229 & 0.762 & Broadened H$\beta$                             \\
G09\_Y1\_DS2\_065 & 0.225 & Extra complex structure around Balmer lines \\
G12\_Y1\_FN1\_127 & 0.270 & Extra complex structure around Balmer lines \\
G12\_Y1\_IS2\_153 & 0.103 & Extra structure at Balmer lines          \\
G09\_Y4\_215\_016 & 0.284 & Extremely broadened Balmer lines         \\
G12\_Y1\_AT\_117  & 0.135 & Extremely broadened Balmer lines and fringing \\
G09\_Y4\_249\_125 & 0.260 & Extremely broadened H$\alpha$                   \\
G15\_Y6\_083\_065 & 1.345 & High redshift quasar: CIII and extra structure \\
G09\_Y3\_015\_281 & 1.128 & High redshift quasar: CIII and MgII      \\
G12\_Y3\_006\_046 & 1.535 & High redshift quasar: CIII and MgII      \\
G09\_Y2\_017\_284 & 0.906 & High redshift quasar: CIII and MgII  \\
G15\_Y4\_212\_020 & 1.885 & High redshift quasar: CIV high and sharp \\
G15\_Y5\_035\_226 & 2.031 & High redshift quasar: CIV high and sharp \\
G15\_Y6\_094\_029 & 1.995 & High redshift quasar: CIV high and sharp \\
G09\_Y4\_226\_135 & 2.563 & High redshift quasar: Lyman-$\alpha$ and broad C \\
G12\_Y6\_073\_125 & 2.232 & High redshift quasar: Lyman-$\alpha$ and Silicon \\
G09\_Y4\_201\_018 & 2.102 & High redshift quasar: Lyman-$\alpha$ emitter \\
G09\_Y4\_212\_196 & 2.238 & High redshift quasar: Lyman-$\alpha$ emitter \\
G15\_Y6\_086\_131 & 3.059 & High redshift quasar: Lyman-$\alpha$ emitter \\
G15\_Y4\_203\_235 & 2.932 & High redshift quasar: Lyman-$\gamma$ emitter \\
G15\_Y5\_021\_245 & 2.837 & High redshift quasar: Lyman-$\gamma$ emitter \\
\hline
\caption{Quasi-Stellar Objects. These objects are either at high redshift of show activity via extremely broadened emission lines.}
\label{tab:qso}
\end{tabularx}

\clearpage

\tiny
\centering
\begin{tabularx}{\linewidth}{l|c|X|l|c|X}
\hline
SPECID & z & Comments \hspace{5cm} & SPECID  & z & Comments \hspace{5cm} \\
\hline
G02\_Y4\_021\_114 & 0.006 &   &G15\_Y1\_GS1\_095 & 0.180 &   \\
G02\_Y4\_029\_209 & 0.012 &   &G15\_Y1\_GX2\_192 & 0.027 &   \\
G02\_Y4\_034\_330 & 0.056 &   &G15\_Y1\_IN1\_030 & 0.035 &   \\
G02\_Y4\_040\_299 & 0.026 &   &G15\_Y1\_IN2\_106 & 0.119 &   \\
G02\_Y4\_041\_266 & 0.183 &   &G15\_Y2\_001\_254 & 0.030 &   \\
G02\_Y4\_043\_105 & 0.078 &   &G15\_Y2\_013\_055 & 0.041 &   \\
G02\_Y4\_043\_105 & 0.078 &   &G15\_Y2\_022\_031 & 0.049 &   \\
G02\_Y5\_079\_023 & 0.015 &   &G15\_Y3\_005\_049 & 0.176 &   \\
G02\_Y5\_082\_208 & 0.073 &   &G15\_Y3\_005\_049 & 0.176 &   \\
G02\_Y5\_091\_065 & 0.078 &   &G15\_Y3\_014\_270 & 0.086 &   \\
G02\_Y5\_114\_257 & 0.083 &   &G15\_Y3\_021\_322 & 0.130 &   \\
G02\_Y5\_114\_291 & 0.166 &   &G15\_Y3\_037\_237 & 0.156 &   \\
G02\_Y6\_004\_067 & 0.297 &   &G15\_Y3\_050\_003 & 0.244 &   \\
G09\_Y1\_AS2\_047 & 0.077 &   &G15\_Y4\_203\_262 & 0.087 &   \\
G09\_Y1\_BS1\_271 & 0.052 &   &G15\_Y4\_214\_347 & 0.045 &   \\
G09\_Y1\_CX2\_244 & 0.028 &   &G15\_Y4\_216\_266 & 0.061 &   \\
G09\_Y1\_EN2\_104 & 0.013 &   &G15\_Y4\_224\_130 & 0.063 &   \\
G09\_Y1\_EN2\_256 & 0.035 &   &G15\_Y4\_234\_096 & 0.185 &   \\
G09\_Y1\_ES2\_147 & 0.070 &   &G15\_Y5\_003\_334 & 0.084 &   \\
G09\_Y1\_FS1\_037 & 0.123 &   &G15\_Y6\_073\_241 & 0.058 &   \\
G09\_Y1\_FS1\_089 & 0.028 &   &G15\_Y6\_073\_391 & 0.073 &   \\
G09\_Y1\_FS2\_387 & 0.018 &   &G15\_Y6\_075\_075 & 0.139 &   \\
G09\_Y1\_IS2\_176 & 0.076 &   &G15\_Y6\_088\_096 & 0.048 &   \\
G09\_Y2\_042\_055 & 0.024 &   &G15\_Y6\_088\_271 & 0.107 &   \\
G09\_Y4\_205\_318 & 0.123 &   &G15\_Y6\_097\_155 & 0.042 &   \\
G09\_Y4\_207\_223 & 0.332 &   &G15\_Y6\_099\_116 & 0.369 &   \\
G09\_Y4\_215\_073 & 0.065 &   &G15\_Y6\_100\_125 & 0.032 &   \\
G09\_Y4\_251\_217 & 0.012 &   &G15\_Y6\_100\_221 & 0.025 &   \\
G12\_Y1\_AN1\_102 & 0.027 &   &G02\_Y3\_018\_304 & 0.139 & BPT Outlier \\
G12\_Y1\_AN1\_392 & 0.018 &   &G02\_Y4\_002\_108 & 0.005 & BPT Outlier \\
G12\_Y1\_AS2\_046 & 0.118 &   &G02\_Y4\_042\_264 & 0.007 & BPT Outlier \\
G12\_Y1\_CN1\_078 & 0.006 &   &G02\_Y5\_061\_198 & 0.013 & BPT Outlier \\
G12\_Y1\_CN1\_102 & 0.040 &   &G09\_Y1\_CN1\_168 & 0.132 & BPT Outlier \\
G12\_Y1\_CND1\_285 & 0.103 &   &G09\_Y1\_DX1\_117 & 0.008 & BPT Outlier \\
G12\_Y1\_CS2\_163 & 0.013 &   &G09\_Y4\_207\_303 & 0.246 & BPT Outlier \\
G12\_Y1\_DND1\_283 & 0.011 &   &G09\_Y4\_207\_339 & 0.157 & BPT Outlier \\
G12\_Y1\_DX1\_211 & 0.140 &   &G09\_Y4\_207\_347 & 0.190 & BPT Outlier \\
G12\_Y1\_FX1\_031 & 0.080 &   &G09\_Y4\_207\_378 & 0.273 & BPT Outlier \\
G12\_Y1\_GN1\_263 & 0.044 &   &G09\_Y4\_231\_266 & 0.012 & BPT Outlier \\
G12\_Y1\_HS2\_086 & 0.190 &   &G12\_Y1\_AN1\_124 & 0.076 & BPT Outlier \\
G12\_Y1\_IS1\_301 & 0.021 &   &G12\_Y1\_AN1\_254 & 0.004 & BPT Outlier \\
G12\_Y1\_IS2\_193 & 0.088 &   &G12\_Y1\_AS2\_119 & 0.006 & BPT Outlier \\
G12\_Y1\_ND1\_188 & 0.027 &   &G12\_Y1\_BS1\_058 & 0.072 & BPT Outlier \\
G12\_Y3\_014\_242 & 0.179 &   &G12\_Y1\_CN1\_079 & 0.013 & BPT Outlier \\
G12\_Y4\_210\_280 & 0.249 &   &G12\_Y1\_IS1\_078 & 0.006 & BPT Outlier \\
G12\_Y6\_056\_022 & 0.312 &   &G12\_Y1\_IS2\_123 & 0.008 & BPT Outlier \\
G12\_Y6\_060\_295 & 0.049 &   &G12\_Y1\_ND8\_039 & 0.022 & BPT Outlier \\
G15\_Y1\_BS1\_108 & 0.066 &   &G12\_Y6\_050\_014 & 0.004 & BPT Outlier \\
G15\_Y1\_CN1\_059 & 0.052 &   &G15\_Y1\_CX1\_053 & 0.030 & BPT Outlier \\
G15\_Y1\_CN2\_291 & 0.116 &   &G15\_Y1\_GN1\_001 & 0.032 & BPT Outlier \\
G15\_Y1\_CS1\_112 & 0.133 &   &G15\_Y3\_009\_097 & 0.026 & BPT Outlier \\
G15\_Y1\_CS1\_149 & 0.051 &   &G15\_Y3\_016\_128 & 0.033 & BPT Outlier \\
G15\_Y1\_CX1\_125 & 0.091 &   &G15\_Y3\_047\_035 & 0.037 & BPT Outlier \\
G15\_Y1\_DS2\_286 & 0.029 &   &G15\_Y4\_205\_288 & 0.007 & BPT Outlier \\
G15\_Y1\_FS1\_068 & 0.103 &   &G15\_Y4\_234\_172 & 0.035 & BPT Outlier \\
G15\_Y1\_FS1\_102 & 0.037 &   &G15\_Y6\_088\_244 & 0.038 & BPT Outlier \\
G15\_Y1\_FS2\_134 & 0.082 &   &G15\_Y6\_089\_265 & 0.099 & BPT Outlier \\
\hline
\caption{Star-Formation Galaxies. Objects with at least very high and narrow Balmer lines. A few of these objects are also BPT outliers as commented on in Section \ref{sec:outlier}.}
\label{tab:sf}
\end{tabularx}

\tiny
\centering
\begin{tabularx}{\linewidth}{l|c|X}
\hline
SPECID & z & Comments \hspace{11cm} \\
\hline
G09\_Y4\_240\_259 & 0.084 & Dominating NII                           \\
G02\_Y3\_013\_192 & 0.070 & Dominating NII, no H$\alpha$                    \\
G09\_Y1\_CN2\_290 & 0.209 & High emission lines and Iron lines       \\
G02\_Y3\_016\_160 & 0.055 & Low H$\alpha$ and no Oxygen                     \\
G12\_Y1\_GND1\_125 & 0.050 & Mysterious emission line at 5435\angstrom         \\
G12\_Y1\_HND1\_169 & 0.199 & NII 5200 emission and other emission lines  \\
G02\_Y3\_014\_029 & 0.056 & NII much bigger than H$\alpha$                  \\
G09\_Y2\_016\_087 & 0.052 & Only H$\alpha$/NII structure                     \\
G12\_Y1\_CX1\_394 & 0.132 & Only H$\alpha$/NII structure                     \\
G12\_Y1\_DS2\_363 & 0.132 & Only H$\alpha$/NII structure                     \\
G12\_Y6\_052\_104 & 0.038 & Only H$\alpha$/NII structure                     \\
G15\_Y1\_FS2\_063 & 0.081 & Only H$\alpha$/NII structure                     \\
G15\_Y4\_204\_298 & 0.136 & Only H$\alpha$/NII structure                     \\
G09\_Y4\_207\_090 & 0.245 & Only H$\alpha$/NII structure                     \\
G09\_Y6\_093\_017 & 0.019 & Only H$\alpha$/NII structure                     \\
G09\_Y4\_207\_260 & 0.246 & Only H$\alpha$/NII structure                    \\
G15\_Y1\_CS2\_180 & 0.139 & Additional Iron line at 5304\angstrom                \\
\hline
\caption{Galaxies grouped on unusual extra lines or weird line fluxes}
\label{tab:lines}
\end{tabularx}

\clearpage

\tiny
\centering
\begin{tabularx}{\linewidth}{l|c|X}
\hline
SPECID & z & Comments \hspace{11cm} \\
\hline
G09\_Y4\_215\_048 & 0.194 &                 \\
G09\_Y4\_215\_057 & 0.238 &                 \\
G12\_Y1\_AN1\_085 & 0.294 &                 \\
G12\_Y1\_CS1\_087 & 0.084 & Minor absorption in H$\delta$ and H$\alpha$ emission          \\
G12\_Y2\_041\_076 & 0.133 &    \\
G15\_Y4\_210\_008 & 0.186 & H$\alpha$ emission       \\
G09\_Y1\_FN2\_235 & 0.293 &                \\
G12\_Y1\_CN1\_175 & 0.177 & H$\alpha$ emission \\
\hline
\caption{Spectra showing H$\delta$ absorption}
\label{tab:ea}
\end{tabularx}

\tiny
\centering
\begin{tabularx}{\linewidth}{l|c|X}
\hline
SPECID & z & Comments \hspace{11cm} \\
\hline
G09\_Y1\_IS2\_022 & 0.055 & Blue galaxy                              \\
G15\_Y1\_DS2\_125 & 0.086 & Blue galaxy                              \\
G15\_Y4\_211\_119 & 0.068 & Blue galaxy                              \\
G12\_Y1\_BD1\_108 & 0.264 & Blue galaxy                              \\
G15\_Y1\_CS1\_071 & 0.085 & Blue galaxy                              \\
G15\_Y1\_DS2\_102 & 0.138 & Blue galaxy                              \\
G02\_Y4\_043\_333 & 0.206 & Diagonal blue end with sky lines in red  \\
G09\_Y4\_211\_013 & 0.186 & Diagonal red                             \\
G09\_Y5\_015\_117 & 0.035 & Diagonal red                             \\
G12\_Y1\_DS2\_367 & 0.165 & Diagonal red                             \\
G02\_Y3\_018\_233 & 0.087 & Diagonal red, no emission lines but strong NaID \\
G09\_Y4\_207\_387 & 0.405 & Diagonal red, no emission lines but strong NaID \\
G12\_Y4\_205\_395 & 0.238 & Diagonal up to 7000A and then flat       \\
G02\_Y4\_043\_265 & 0.260 & High blue end                            \\
G09\_Y4\_215\_246 & 0.018 & High blue end                            \\
G15\_Y1\_DS2\_097 & 0.053 & High blue end                            \\
G15\_Y1\_DS2\_106 & 0.193 & High blue end                            \\
G15\_Y1\_DS2\_123 & 0.179 & High blue end                            \\
G15\_Y1\_DS2\_124 & 0.053 & High blue end                            \\
G15\_Y4\_202\_230 & 0.204 & High blue end                            \\
G15\_Y4\_202\_266 & 0.099 & High blue end                            \\
G15\_Y4\_202\_314 & 0.025 & High blue end                            \\
G15\_Y4\_204\_220 & 0.085 & High blue end                            \\
G15\_Y6\_074\_314 & 0.027 & High blue end                            \\
G15\_Y6\_088\_239 & 0.025 & High blue end                            \\
G15\_Y4\_237\_091 & 0.224 & Mountain shape                           \\
G09\_Y1\_DS2\_261 & 0.108 & Noisy blue end                           \\
G15\_Y4\_204\_386 & 0.122 & Noisy blue end                           \\
G15\_Y4\_204\_396 & 0.053 & Noisy blue end                           \\
G02\_Y3\_019\_369 & 0.323 & Valley shape                             \\
G09\_Y4\_212\_315 & 0.079 & Valley shape                             \\
G09\_Y4\_212\_321 & 0.040 & Valley shape                             \\
G15\_Y6\_082\_045 & 0.025 & Very weird feature                       \\
\hline
\caption{Outlying spectra based on their continuum. These are either extremely blue or diagonal red galaxies due to clustering of the URF algorithm, or weird features on a continuum level.}
\label{tab:continuum}
\end{tabularx}

\tiny
\centering
\begin{tabularx}{\linewidth}{l|c|X}
\hline
SPECID & z & Comments \hspace{11cm} \\
\hline
G15\_Y2\_009\_099 & 0.256 & Blue end of spectrum diagonal            \\
G12\_Y1\_HS2\_140 & 0.074 & Dominating spurious line                 \\
G12\_Y1\_CND1\_092 & 0.060 & Noisy continuum                          \\
G15\_Y2\_020\_394 & 0.179 & Reduction error with sky line 5578       \\
G15\_Y4\_215\_093 & 0.312 & Sky absorption at 5578                   \\
G15\_Y6\_081\_395 & 0.052 & Sky absorption at 5578                   \\
G02\_Y5\_079\_376 & 0.142 & Sky lines not reduced correctly          \\
G12\_Y1\_AS2\_129 & 0.076 & Sky lines not reduced correctly          \\
G12\_Y1\_GS1\_117 & 0.186 & Sky lines not reduced correctly          \\
G15\_Y2\_008\_251 & 0.145 & Sky lines not reduced correctly          \\
G15\_Y4\_207\_028 & 0.249 & Spectral arms peak at splice             \\
G09\_Y1\_AN1\_378 & 0.201 & Very dominant sky absorption             \\
G09\_Y1\_AX2\_247 & 0.190 & Very dominant sky absorption             \\
G09\_Y1\_CN1\_233 & 0.166 & Very dominant sky absorption             \\
G15\_Y1\_BS2\_216 & 0.135 & Very dominant sky absorption             \\
G02\_Y5\_060\_294 & 0.013 & Weird absorption feature (all fibre 294?) \\
G15\_Y2\_009\_294 & 0.166 & Weird absorption feature (all fibre 294?) \\
G15\_Y2\_020\_294 & 0.208 & Weird absorption feature (all fibre 294?) \\
G15\_Y3\_051\_294 & 0.144 & Weird absorption feature (all fibre 294?) \\
\hline
\caption{Spectra containing reduction errors or other features not belonging to the objects}
\label{tab:reduction}
\end{tabularx}
\end{document}